\documentclass[twocolumn,aps,prb,superscriptaddress,a4paper,floatfix]{revtex4-1}

\usepackage{graphicx}% Include figure files
\usepackage{dcolumn}% Align table columns on decimal point
\usepackage{bm}% bold math
\usepackage{type1cm}% scalable Adobe fonts only
\usepackage[normalem]{ulem}

\usepackage{amssymb,amsmath}

\usepackage[dvips]{color}

\newcommand{\mm}[1]{\mbox{$#1$}}
\newcommand{\dstd}{\mathrm{d}}

\newcommand{\msp}{\mbox{$\mu_{\mathrm{spin}}$}}
\newcommand{\mo}{\mbox{$\mu_{\mathrm{orb}}$}}
\newcommand{\mB}{\mbox{$\mu_{B}$}}

\newcommand{\Bex}{\mbox{$B_{\text{ex}}$}}
\newcommand{\mae}{\mbox{$E_{\text{MAE}}$}}

\newcommand{\ea}{{\it et al.}}

\begin{document}

\title{Magnetocrystalline anisotropy energy for adatoms and monolayers
  on non-magnetic substrates: where does it comes from?}

% repeat the \author .. \affiliation  etc. as needed
% \email, \thanks, \homepage, \altaffiliation all apply to the current
% author. Explanatory text should go in the []'s, actual e-mail
% address or url should go in the {}'s for \email and \homepage.
% Please use the appropriate macro for each type of information

% \affiliation command applies to all authors since the last
% \affiliation command. The \affiliation command should follow the
% other information
% \affiliation can be followed by \email, \homepage, \thanks as well.

\author{O. \surname{\v{S}ipr}} 
\email{sipr@fzu.cz}
\homepage{http://www.fzu.cz/~sipr} \affiliation{Institute of Physics
  of the ASCR v.~v.~i., Cukrovarnick\'{a}~10, CZ-162~53~Prague, Czech
  Republic }

\author{S. \surname{Bornemann}} \affiliation{Universit\"{a}t
  M\"{u}nchen, Department Chemie, Butenandtstr.~5-13,
  D-81377~M\"{u}nchen, Germany}

\author{H. \surname{Ebert}} \affiliation{Universit\"{a}t M\"{u}nchen,
  Department Chemie, Butenandtstr.~5-13, D-81377~M\"{u}nchen, Germany}

\author{J. \surname{Min\'{a}r}} \affiliation{Universit\"{a}t
  M\"{u}nchen, Department Chemie, Butenandtstr.~5-13,
  D-81377~M\"{u}nchen, Germany}

%%%\date{\today}

\begin{abstract}
The substrate contribution to the magnetic anisotropy energy (MAE) of
supported nanostructures can be assessed by a site-selective
manipulation of the spin-orbit coupling (SOC) and of the effective
exchange field $\Bex$.  A systematic study of Co adatoms and
Co monolayers on the (111) surfaces of Cu, Ag, Au, Pd and Pt is
performed to study common trends in this class of materials.  It is
found that for adatoms, the 
influence of the substrate SOC and \Bex\ 
is relatively small
(10--30\% of the MAE) while for monolayers, this influence
can be substantial.  The influence of the substrate SOC is much more
important than the influence of the substrate 
$\Bex$, except for highly polarizable substrates with a
strong SOC (such as Pt).  
The substrate always promotes the tendency to an out-of-plane
  orientation of the easy magnetic axis for all the 
  investigated systems.
\end{abstract}

\pacs{75.30.Gw,75.70.Tj,75.70.Ak}

\keywords{magnetism,anisotropy,spin-orbit coupling}

\maketitle

%%%%%%%%%%%%%%%%%%%%%%%%%%%%%%%%%%%%%%%%%%%%%%%%%%%%%%%%%%%%%%%

\section{Introduction}

One of the areas of materials research where a lot of effort has
concentrated lately is artificially prepared systems composed of
magnetic and non-magnetic elements.  This includes multilayers, thin
films, monolayers and various nanostructures supported by substrates.
One of the properties in focus here is the magnetocrystalline
anisotropy, i.e., the tendency of the system to orient its
magnetization preferentially in one direction with respect to the
crystal lattice.  Such a property is very appealing for technology as
it finds its application in device and information technology and in
the whole area of what is now called spintronics.  However,
magnetocrystalline anisotropy presents also an interesting topic for
fundamental physics.

{\em Ab initio} calculations of the magnetic anisotropy energy (MAE)
are especially challenging.  First, there are big technical
difficulties as evaluating the MAE means that, at least in principle,
one has to subtract two large numbers --- total energies for two
orientations of the magnetization --- with a great accuracy.  Second,
the calculated value of the MAE can be affected by various factors
that are hard to control, such as many-body effects beyond the local
density approximation (LDA) \cite{YSK+01,SM+03,STK+05} or ``boundary
effects'' due to the finite sizes of the supercells used for
describing the nanostructure \cite{SBME10}. However, apart from
calculating the MAE as accurately as possible one also wants to get an
intuitive feeling how it arises.  One direction on this front is to
try to link the MAE to changes of specific electron
states \cite{WWF93a,DKS+94,GC+12}. Another direction, pursued strongly when
studying composed systems, is to try to answer the question: ``where
does the MAE come from''?  Which atoms contribute to the MAE to which
extent?  What is the role of the substrate for the MAE of adatoms and
monolayers?  What is the role of non-magnetic atoms in layered
compounds such CoPd or FePt?

%%%------------------%%%------------------%%%---------------%%%

\subsection{Assigning the MAE to individual atoms}

\label{sec-uptonow}

There are several ways how one could try to assign MAE contributions
to individual atoms.  Probably the one that comes most handy is to
make use of the fact that the expression for the MAE usually contains
a sum over atomic sites in one way or another.  If the MAE is
calculated directly as a difference between total energies for two
orientations of the magnetization, one can subdivide the spatial
integral which has to be evaluated in this case into parts coming from
different regions.  This approach was adopted by Tsujikawa and
Oda \cite{TO+09} to study the spatial variations of the MAE for
Pt/Fe/Pt(001) and FePt.  A similar philosophy can be applied also if
the MAE is calculated simply by subtracting the band energies, relying
on the magnetic force theorem.  There again one can identify
contributions from different atoms because the expression contains
site-projected densities of states.  
In this way the localization of the MAE was studied, e.g., for
  YCo$_{4}$ \cite{NBL+92}, Fe and Co thin films on
  Cu \cite{USW96,HUZ+02}, Co monolayer on Au \cite{USBW96}, Co/Pd
  structures \cite{CE+97,DDP03}, FePt and Fe$_{1-x}$Mn$_{x}$Pt
  alloys \cite{BES+05}, Fe and Co wires on Pt \cite{KSF+06} or Fe and Co
  adatoms on Rh and Pd \cite{BLD+10}.

Another appealing possibility is to make use of the torque formula
(which again relies on the magnetic force theorem).  Here one formally
adds contributions originating from the torque exerted on individual
localized magnetic moments, so one can presume that these quantities
correspond to real physical quantities (even though these moments are
not independent, because the torque formula assumes that {\em all} the
moments are infinitesimally rotated at the same time).  Materials
which were investigated in this way recently include Fe, Co and Ni
adatoms and monolayers on Ir, Pt and Au \cite{SBME10,BSM+12}, or a
diluted Pt monolayer inside Co \cite{APS+12}. 

Apart from making use of formal sums over atomic sites which occur in
the expressions used for evaluating the MAE, one can turn to various
models and try to get an insight from there.  In this respect the
so-called Bruno formula, linking the MAE to the anisotropy of the
orbital magnetic moment \mo, comes into mind: one can define that for
a multicomponent system, the relative importance of different atomic
types is proportional to the anisotropies of \mo\ for those types.
This approach was applied, e.g., to metallic films \cite{DVS+08} or
Co/Ni (111) superlattices \cite{GC+12}.

There is yet another possibility to use a model-based approach to
identify localized contributions to the MAE.  Namely, as the
magnetocrystalline anisotropy cannot arise without the spin-orbit
coupling (SOC), one can think of introducing a non-zero SOC only at
some atomic sites while keeping it zero at the remaining ones.  The
MAE of such a model system could then be seen as the MAE due to those
atoms where the SOC has been kept.  Such an approach was used, e.g.,
by Wang \ea \cite{WWF93a} for a Pd/Co/Pd sandwich, \'{U}jfalussy
\ea \cite{USW97} for Fe/Cu thin film overlayers, Baud \ea \cite{BRB+06}
for Co wires on Pt or Subkow and F\"{a}hnle \cite{SF+11} for Fe films.
 A similar approach can be adopted in another formulation, where
  the torque is evaluated as a sum of terms due to the SOC at
  different atomic types \cite{KSM+11}.  In this way contributions to
  MAE were analyzed, e.g., for a Mn monolayer on W(001) \cite{SMO+08}
  or for antiferromagnetic MnX layered materials (X = Ni, Pd, Rh, Ir)
  \cite{KSM+11}.
Also other authors relied on SOC manipulation to demonstrate the
importance of non-magnetic atoms for the MAE of layered compounds of
3$d$ and 4$d$/5$d$ noble
metals \cite{DKS+94,SDM95,RKF+01,Cin+01,BES+05}, even though they did
not perform a full analysis of the various contributions.  

By analyzing the results obtained for various systems up to now
(mostly but not exclusively via the torque decomposition scheme), a
prevailing pattern concerning the importance of non-magnetic element
for the MAE emerges: (i)~For 3$d$ adatoms on non-magnetic substrates,
the contribution from the magnetic atom clearly dominates,
the contribution from the non-magnetic atoms can be
neglected \cite{SBME10,BSM+12,SML+09}. (ii)~For monolayers or wires of
3$d$ atoms on non-magnetic substrates as well as for layered systems
such as L1$_{0}$ compounds, the role of the substrate is important,
sometimes even
dominant \cite{BES+05,BRB+06,KSF+06,SMO+08,SBME10,KSM+11,BSM+12}.

%%%------------------%%%------------------%%%---------------%%%

\subsection{Problems with assigning MAE to individual atoms}

Despite the common trends that can be extracted from the results
obtained via various methods, there are clearly also problems with
attempts to attribute the MAE to individual atoms.  First, one should
mention that even technically, the decomposition schemes may be
ambiguous, as it was demonstrated by Subkow and F\"{a}hnle \cite{SF+09}
for a particular implementation of the decomposition via the sum over
the band energies.  Second, if different MAE decomposition schemes
are applied to the same system, different results are obtained.  For
example, Burkert \ea \cite{BES+05} investigated FePt and found that Fe
atoms are responsible for about 70\% of the MAE if estimated from the
sum of the band energies but only for 14\% of the MAE if estimated by
manipulating with the SOC.  This controversy can be reinforced by
observing that if the MAE is decomposed via subdivision of the
spatial integral in the total energy formula, the contribution from Fe
atoms is more important than the contribution from Pt
atoms \cite{TO+09} while if the decomposition via real-space
calculation of the torque is employed, the MAE is attributed almost
entirely to the Pt sublattice \cite{SDM95}.

While one might be able to reconcile these differences by one way or
another, there is a deep internal problem with the effort to assign
MAE contributions to individual atoms.  Namely, 
total energy of an inhomogeneous
system is not an extensive quantity and so one cannot decompose the
MAE uniquely into sums of contributions coming from various spatial
regions.  Assigning one part of the MAE to one atom and another part
of the MAE to a different atom can be always only intuitive and in
principle ambiguous.  It is only the final sum that provides a
well-defined quantity.

This does not necessarily mean that there can never be any physical
content in the division of the MAE among atoms or layers.  E.g., it
was found that if the MAE for a Co layer buried in Pt is decomposed
via the torque formula, the layer-resolved MAE obtained in this way
can be related to the shifts of the valence states in a given layer
invoked by Friedel oscillations of the charge \cite{APS+12}, which
clearly is a well defined physical concept.  However, in general,
whenever one tries to decompose the MAE into a sum of spatially
located parts, one has to be prepared for ambiguities and
inconsistencies.

%%%------------------%%%------------------%%%---------------%%%

\subsection{Method for assessing the role of individual sites for the MAE}

\label{sec-decomp}

Despite all this, one would still like to know what is the role of
different atoms for the MAE of complex systems, even though such a
question has formally no exact meaning; the intuitive meaning of such
a question is clear enough.  One just needs to reformulate the
question about the localization of the MAE in such a way that it is
formally well-defined and yet it embraces the vague but illuminating
concept of ``where does the MAE come from''.

To find such a formulation, one should take into account what is the
mechanism that leads to the magnetocrystalline anisotropy --- in
particular, in systems comprising magnetic and non-magnetic atoms
alike.  Let us recall that Shick \ea \cite{SMO+08}, suggested that in
3$d$--5$d$ bimetallic systems such as CoPt and FePt, the contribution
of the 5$d$ atoms to the MAE originates from (i)~strong SOC at the
5$d$ atoms, (ii)~exchange splitting at the 5$d$ atoms induced by the
magnetic 3$d$ atoms, and (iii)~Stoner enhancement of the local spin
susceptibility at 5$d$ atoms.  Items (i) and (iii) can be tested by a
computer experiment: one can selectively switch off the SOC at
non-magnetic atoms via various schemes and one can also suppress the
effective exchange field \Bex\ at non-magnetic atoms by simply forcing
the spin-up and spin-down potentials to be equal.  In this way one
gets a well-defined model system for which a uniquely defined
MAE can be calculated.  As concerns the exchange splitting induced by
the magnetic atoms at the originally non-magnetic atoms [the point
  (ii) mentioned above], this cannot be completely eliminated:  by
suppressing the \Bex\ field for non-magnetic atoms, one does not
remove the magnetization of the respective atoms completely but allows
them to be polarized only by hybridization with neighbouring magnetic
atoms.  The local Pauli or paramagnetic susceptibility can be seen as
a measure for the effectiveness of this mechanism.

The MAE obtained when suppressing the SOC and \Bex\ in the region
occupied by originally non-magnetic atoms can be viewed as that part
of the MAE of the system which comes only from the magnetic atoms ---
because the SOC and \Bex\ at the non-magnetic atoms have been
suppressed.  Promoting this picture further, one can think of the
difference between the MAE for the system with full SOC and \Bex\ and
the MAE for the system with SOC and \Bex\ suppressed in the substrate
as of the ``contribution of the substrate'' to the total MAE of the
system.  

Obviously, this is only an intuitive concept that cannot be taken too
literally.  Due to the non-locality of the MAE, any attempt to
decompose it is in principle ambiguous.  E.g., using the same
philosophy as above but along a different path, one could
alternatively define the
contribution of the substrate as the MAE calculated with the SOC
suppressed at the magnetic adatoms.  Such a quantity would clearly
differ from the difference discussed above (see also the results in
Tabs.~\ref{tab-mae-ada} and~\ref{tab-mae-lay} in
Sec.~\ref{sec-role-mae}).  There is no formally exact way of saying
which approach is better than another one.  Still, some approaches may
be preferable in the {\em intuitive} way, by illustrating certain
physical aspects.  Our concept respects the fact that there would be
no MAE without the magnetic adatoms and contains an aspect of spacial
localization via site-related SOC and \Bex.  At the same time, unlike
some other approaches, is it {\em technically} unambiguous because it
always involves calculating the MAE for the whole system.  One can
also view the results presented here disregarding any discussion about
``localization'', simply as a comprehensive study of separate effects
of SOC and \Bex\ on the MAE of adatoms and monolayers.
One should also have in mind that we focus here specifically on the
role of the substrate SOC and \Bex.  However, even without any SOC
and \Bex\ the substrate affects the electron states via hybridization
and thus has an influence on the MAE. This aspect was thoroughly
explored, e.g., when comparing the MAE for a free-standing Co
monolayer and for Co/Cu, Co/Ag, and Co/Pd
multilayers \cite{WWF93a,DKS+94} or for a Co monolayer on
Pt \cite{MLR+08}. In particular, it was noted that the position of the
$d$ states of the magnetic atoms with respect to the $d$ states of the
substrate and to the Fermi energy is important \cite{WWF93a}. In this
work we mean by contribution of the substrate to the MAE only the
contribution of the substrate SOC and \Bex, which can be localized
within the limitations and ambiguities mentioned above.

Our approach towards assessing the role of non-magnetic atoms for the
MAE is in line with earlier works where the SOC was manipulated in a
similar
way \cite{WWF93a,BRB+06,SF+11,DKS+94,SDM95,RKF+01,Cin+01,BES+05}. It
should be noted, however, that despite the numerous works where the
SOC manipulation was used to analyze the MAE, the results obtained so
far are quite sparse and scattered among different systems and it is
hardly possible to draw systematic conclusions concerning the
influence of the SOC at non-magnetic atoms.  The role of the effective
exchange field at originally non-magnetic atoms has not been
investigated in this respect before, to the best of our knowledge.

In this work we want to focus on a series of systems comprising
magnetic adatoms and monolayers on non-magnetic noble metal substrates
and to assess the role of the substrate for the MAE.  We focus
selectively on the role of the SOC and of the effective exchange field
at the substrate atoms.  We will show that while for the adatoms the
 contribution of the substrate SOC and \Bex\ 
is relatively small, for monolayers it can
substantial.  We will also show that the contribution due to the SOC
is more important than the contribution due to \Bex\ and that for
highly polarized substrates with large SOC, the effect of both factors
is non-additive.

%%%%%%%%%%%%%%%%%%%%%%%%%%%%%%%%%%%%%%%%%%%%%%%%%%%%%%%%%%%%%%%
%%%%%%%%%%%%%%%%%%%%%%%%%%%%%%%%%%%%%%%%%%%%%%%%%%%%%%%%%%%%%%%

\section{Computational framework}

% Figure planned for 1 column
\begin{figure}
\includegraphics[viewport=0 0 540 520,height=30mm]{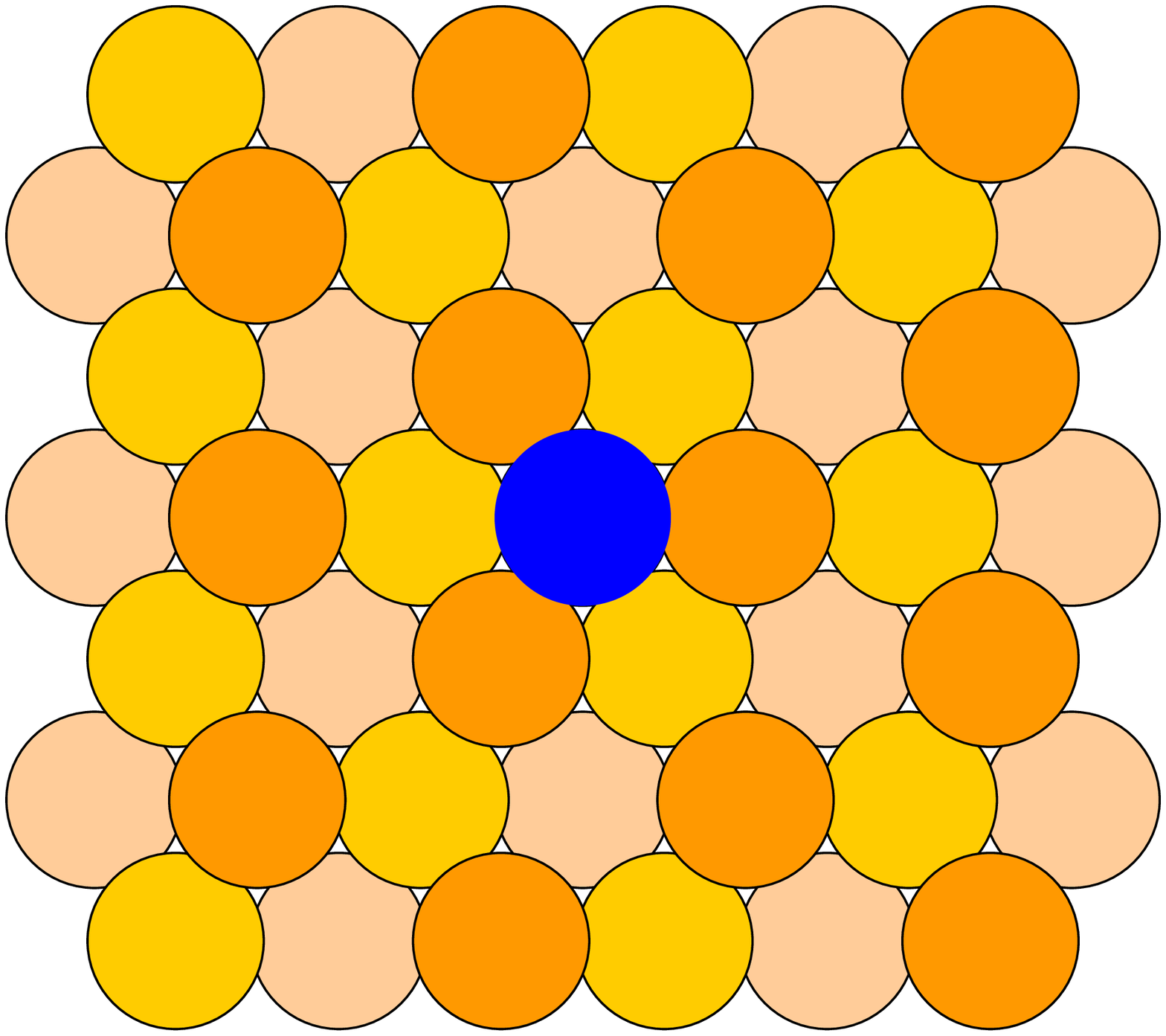}%
\includegraphics[viewport=-160 0 540 520,height=30mm]{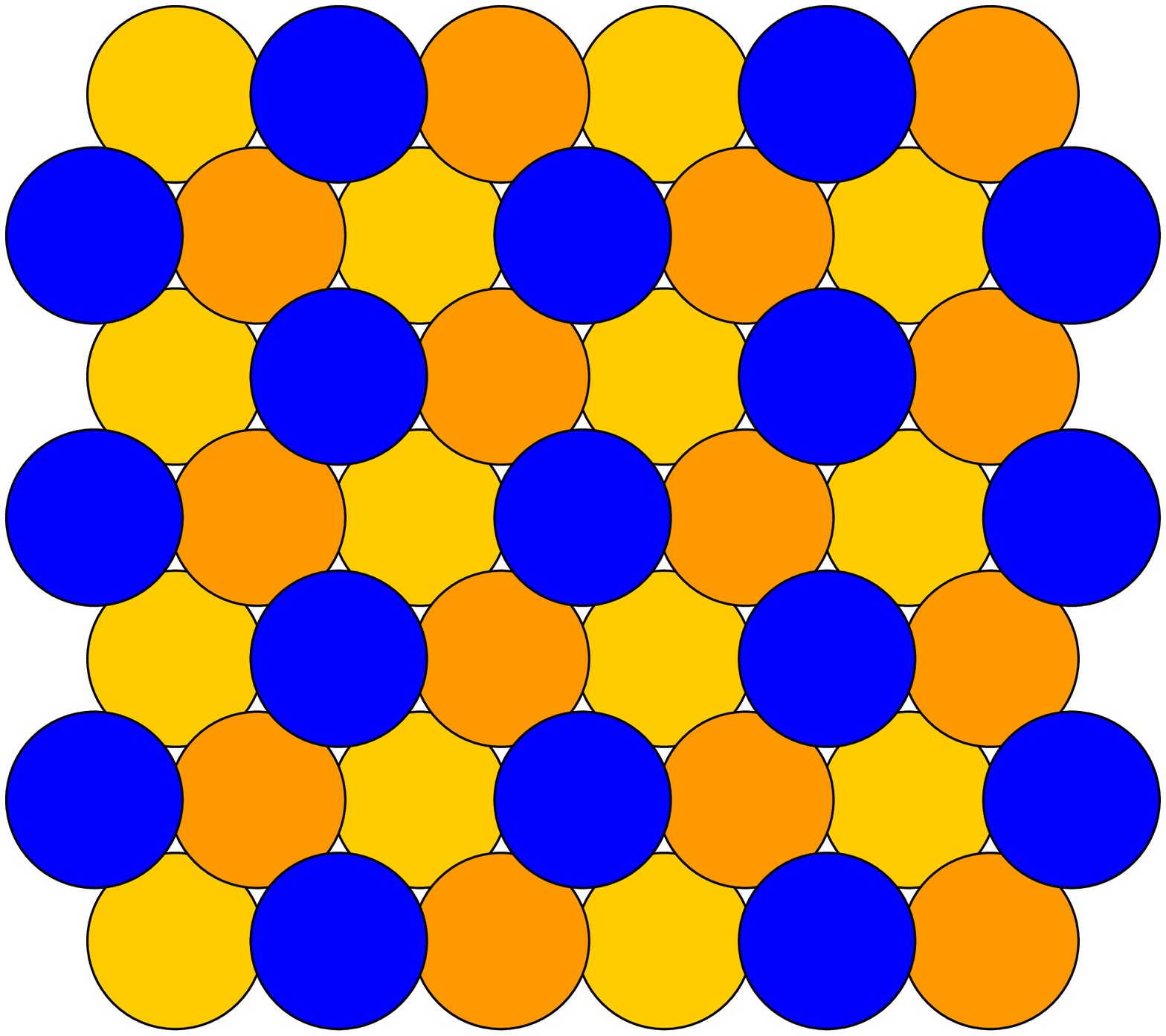}%
\caption{(Color online) Structure diagrams for a Co adatom and a Co
  monolayer on a (111) surface of an fcc crystal. The Co atoms are
  represented by blue (dark) circles, various shades of orange (grey)
  represent substrate atoms in different layers.}
\label{fig-geom}
\end{figure}

We study the MAE for Co adatoms and Co monolayers on (111) surfaces
of noble metals Cu, Ag, Au, Pd, Pt.  In this way we include in our
study substrates which are hard to polarize (Cu, Ag, Au) and
substrates that are easy to polarize (Pd, Pt) as well as substrates
with weak SOC (Cu), with moderate SOC (Pd, Ag), and with strong SOC
(Pt, Au).  The geometry of the systems is schematically shown in
figure~\ref{fig-geom}, some properties of the substrates are summarized
in table~\ref{tab-subs}.  

The electronic structure is calculated within the {\em ab initio} spin
density functional framework, relying on the local spin density
approximation (LSDA) with the Vosko, Wilk and Nusair parametrization
for the exchange and correlation potential \cite{VWN80}. The
electronic structure is described, including all relativistic effects,
by the Dirac equation, which is solved using the spin polarized
relativistic multiple-scattering or Korringa-Kohn-Rostoker (SPR-KKR)
Green's function formalism \cite{EKM11} as implemented in the {\sc
  spr-tb-kkr} code \cite{tbkkr-code}. The potentials were treated
within the atomic sphere approximation (ASA). 
For the multipole expansion of the Green's function, an angular
momentum cutoff \mm{\ell_{\mathrm{max}}}=3 was used.  
 The energy integrals were evaluated by contour integration on a
  semicircular path within the complex energy plane using a
  logarithmic mesh of 32 points.  The integration over the
  $\mathbf{k}$ points was done on a regular mesh, using 10000 points
  in the full surface Brillouin zone. The convergence of the MAE with
  respect to the $\mathbf{k}$ space integration grid is checked
  in~\ref{sec-kgrid}.

% Table planned for a single column
\begin{table}
\caption{Comparison of properties of noble metals used
  as substrates: 
  lattice constant $a$, the SOC parameter $\xi$, magnetic
  susceptibility $\chi_{m}$ and Stoner enhancement factor $S_{xc}$.
  The SOC parameters $\xi$ were calculated for bulk crystals by a
  method described by Davenport \ea \cite{DWW+88}, the remaining
  properties were taken from the literature.}
\label{tab-subs}
\begin{ruledtabular}
\begin{tabular}{lccrc}
   &  
\multicolumn{1}{c}{$a$ [\AA]} & 
\multicolumn{1}{c}{$\xi$ [meV]} & 
\multicolumn{1}{c}{$\chi_{m}$ [10$^{-6}$cm$^{3}$mol$^{-1}$]} &
    \multicolumn{1}{c}{$S_{xc}$} \\
\hline 
%%%
Cu  &  3.615  &  137  &     -5$^{\rm a}$  &  
\multicolumn{1}{c}{ 1.1,$^{\rm b}$ 1.1$^{\rm c}$ } \\
%%%
Ag  &  4.085  &  299  &   -20$^{\rm a}$  &  
\multicolumn{1}{c}{ 1.1,$^{\rm b}$ 1.2$^{\rm c}$ }  \\
%%%
Au  &  4.078  &  845  &   -28$^{\rm a}$  &  
\multicolumn{1}{c}{ 1.1,$^{\rm b}$ 1.1$^{\rm c}$ } \\
%%%
Pd  &  3.891  &  237  &  540$^{\rm a}$  &  
\multicolumn{1}{c}{ 9.9,$^{\rm d}$ 12.1$^{\rm e}$ } \\
%%%
Pt  &  3.911  &  712  &  193$^{\rm a}$  &  
\multicolumn{1}{c}{ 3.7,$^{\rm d}$ 4.2$^{\rm e}$ } \\
%%%
\end{tabular}
\end{ruledtabular}
$^{\rm a}$ {Lide \cite{CRC+07}} \\
$^{\rm b}$ {MacDonald \ea \cite{MDV+82}} \\
$^{\rm c}$ {Smelyansky \ea \cite{SLP+93}} \\
$^{\rm d}$ {S\"{a}nger and Voitl\"{a}nder \cite{SV+80}} \\
$^{\rm e}$ {Povzner \ea \cite{PVF+10}}
\end{table}

The electronic structure of Co monolayers on surfaces was calculated
by means of the tight-binding KKR technique \cite{ZDU+95}.
The substrate was modelled by a slab of 16 layers, the vacuum was
represented by 4 layers of empty sites.  The adatoms were treated as
embedded impurities: first the electronic structure of the host system
(substrate with a clean surface) was calculated and then a Dyson equation for an
embedded impurity cluster was solved \cite{BMP+05}. The impurity
cluster contains 131 sites; this includes a Co atom, 70 substrate
atoms and the rest are empty sites.  We assume that all the atoms are
located on ideal lattice sites of the underlying bulk fcc lattice, no
structural optimization was attempted.  While this affects the
comparison of our data with experiment, 
we do not expect this to have 
 significant
influence on our conclusions concerning the relative role of the
substrate for the MAE of adatoms and monolayers 
 (see also the discussion in Sec.~\ref{sec-diss}).

The MAE is calculated by means of the torque
$T_{\hat{u}}(\hat{n})$ which describes the variation of the energy
if the magnetization direction $\hat{n}$\ is infinitesimally rotated
around an axis $\hat{u}$.  If the expansion of the total energy is 
restricted to the second order in directions cosines as \cite{Bir66}
\begin{eqnarray}
E(\theta, \phi) 
   &= E_0 + K_{2,1}\cos{2\theta}      \nonumber \\
   &  \quad  + K_{2,2}(1-\cos{2\theta})\cos{2\phi}  \nonumber \\
   &  \quad  + K_{2,3}(1-\cos{2\theta})\sin{2\phi}  \nonumber \\
   &  \quad  + K_{2,4}\sin{2\theta}\cos{\phi} \nonumber \\
   &  \quad  + K_{2,5}\sin{2\theta}\sin{\phi}       
\label{MAEcoeff}
\;\;, 
\end{eqnarray}
where $\theta$ is the angle between the surface normal and the
magnetization direction and $\phi$ is the azimuthal angle, 
 the difference in energy between the in-plane and out-of-plane
magnetizations can be obtained just by evaluating the torque for
\mm{\theta=45^{\circ}} \cite{WWW+96}:
\begin{eqnarray}
E(90^{\circ}, \phi) - E(0^{\circ}, \phi)
   &= -2K_{2,1} + 2K_{2,2}\cos{2\phi}  \nonumber \\
   & \qquad + 2K_{2,3}\sin{2\phi}  \nonumber \\
   &= \left. \frac{\partial E(\theta, \phi)}{\partial \theta} 
      \right|_{\theta=45^{\circ}}
\;\;.
\label{MAEvalue}
\end{eqnarray}
The torque itself was calculated
by relying on the magnetic force theorem \cite{SSB+06}. We define the
MAE as
\begin{equation}
E_{\mathrm{MAE}} \; \equiv \; E^{(x)} \, - \, E^{(z)}
\;\;,
\label{maedef}
\end{equation}
where $E^{(\alpha)}$ is the total energy of a system if the
magnetization is parallel to the $\alpha$ axis.  A positive \mae\ thus
implies an out-of-plane magnetic easy axis.  

 It should be noted that by evaluating the MAE according to
  equation (\ref{MAEvalue}) which relies on the expansion
  (\ref{MAEcoeff}), we restrict ourselves to the uniaxial contribution
  to the MAE, neglecting the higher order terms.  As our aim is to
  investigate the basic trends concerning the influence of the
  substrate SOC and \Bex\ on the MAE, this simplification does not
  affect our conclusions. A comparison of results obtained via
  equation (\ref{MAEvalue}) with results obtained via a full torque
  integration is presented in~\ref{sec-torque}.

Apart from the magneto-crystalline contribution to the MAE which
  we focus on, there is also a dipole-dipole contribution to the MAE
  due to the Breit interaction (shape anisotropy) \cite{BMB+12}. The
  shape anisotropy energy is not considered here; it's value for a
  Co monolayer on noble metal (111) surfaces is about
  -0.09~meV~\cite{BSM+12}.

If the spin-orbit coupling is included in the calculation implicitly
via the Dirac equation, it is not possible to study the relation
between the SOC strength and a selected physical quantity in a direct
way --- contrary to schemes where a SOC term can be identified in the
approximate Hamiltonian. Rather, one can vary the speed of light $c$
which, however, modifies all relativistic effects.  It is nevertheless
possible to isolate the bare effect of the SOC by using an approximate
two-component scheme \cite{EFVG96} where the SOC-related term is
identified via relying on a set of approximate radial Dirac equations.
In some respect this approach is an extension of the scheme worked out
by Koelling and Harmon \cite{KH77} and implemented by MacLarren and
Victora \cite{MV94}.  This scheme was used in the past to investigate
the influence of the SOC on 
various properties \cite{EVB99,MEG+01,PEP+05,MKWE13}. In this work
we use this scheme to suppress the SOC selectively either at the
substrate atoms or at the Co atoms while retaining all other
relativistic effects.  We checked that if the SOC is included at all
sites 
 within the approximative scheme \cite{EFVG96}, it yields
  practically identical results to those obtained if the full Dirac
  equation is solved 
(for example, the MAE
obtained for the monolayers using a full Dirac equation is by
0.01--0.02~meV per Co atom larger than the MAE obtained using the
approximative scheme of Ebert \ea\ \cite{EFVG96}).

To assess the role of the Stoner enhancement of the local spin
susceptibility at the substrate atoms, we made yet another series of
calculations, with the effective exchange field \Bex\ set to zero for
the substrate atoms during the self-consistent cycle.  In this way the
substrate atoms will be spin-polarized only due to the unenhanced
Pauli susceptibility.  Suppressing \Bex\ would presumably have little
effect for Cu, Ag, or Au substrates where the Stoner enhancement
factor $S_{xc}$ is small.  However, for Pd and Pt, which are close to
the ferromagnetic instability and the $S_{xc}$ factor is relatively
large, suppressing \Bex\ could affect the outcome significantly (see
Polesya \ea \cite{PMS+10} for a comparison of enhanced and unenhanced
spin magnetic moments in Pd).

%%%%%%%%%%%%%%%%%%%%%%%%%%%%%%%%%%%%%%%%%%%%%%%%%%%%%%%%%%%%%%%
%%%%%%%%%%%%%%%%%%%%%%%%%%%%%%%%%%%%%%%%%%%%%%%%%%%%%%%%%%%%%%%

\section{Results}

%%%------------------%%%------------------%%%---------------%%%

\subsection{Influence of SOC and \Bex\ on magnetic moments}

% Wide table planned for two columns
\begin{table}
\caption{Spin magnetic moments \msp\ (in \mB) inside atomic
  spheres around Co adatoms on noble metals, with magnetization
  perpendicular to the surface.  The first data 
  column shows the results with the SOC included both at the Co atoms
  and at the substrate, the second column shows the results with the
  SOC only at Co atoms, and the last column shows the results with
  the SOC considered only at the substrate atoms.  Additionally, the
  first line for each substrate shows the results obtained if no
  restrictions are laid on the exchange field \Bex\ while the second
  line shows results obtained if \Bex\ is suppressed in the
  substrate.}
\label{tab-spn-ada}
\begin{ruledtabular}
\begin{tabular}{llccc}
   &  & 
    \multicolumn{1}{c}{$\xi_{\mathrm{Co}} \ne 0$} & 
    \multicolumn{1}{c}{$\xi_{\mathrm{Co}} \ne 0$} & 
    \multicolumn{1}{c}{$\xi_{\mathrm{Co}} = 0$} 
  \\
   \multicolumn{2}{l}{substrate} & 
    \multicolumn{1}{c}{$\xi_{\mathrm{sub}} \ne 0$} & 
    \multicolumn{1}{c}{$\xi_{\mathrm{sub}} = 0$} & 
    \multicolumn{1}{c}{$\xi_{\mathrm{sub}} \ne 0$}  
  \\
%%%\hline  \\  [-3.5ex]
\hline
%%%
Cu & $B_{\mathrm{ex}} \ne 0$ & 2.11 & 2.11 & 2.11  \\
   & $B_{\mathrm{ex}} = 0$   & 2.11 & 2.11 & 2.11  \\
  [0.7ex]
%%%                                                        
Ag & $B_{\mathrm{ex}} \ne 0$ & 2.22 & 2.22 & 2.22  \\
   & $B_{\mathrm{ex}} = 0$   & 2.22 & 2.22 & 2.22 \\
  [0.7ex]
%%%                                               
Au & $B_{\mathrm{ex}} \ne 0$ & 2.28 & 2.28 & 2.28  \\
   & $B_{\mathrm{ex}} = 0$   & 2.28 & 2.28 & 2.28  \\
  [0.7ex]
%%%                                               
Pd & $B_{\mathrm{ex}} \ne 0$ & 2.33 & 2.33 & 2.33  \\
   & $B_{\mathrm{ex}} = 0$   & 2.34 & 2.33 & 2.34  \\
  [0.7ex]
%%%                                               
Pt & $B_{\mathrm{ex}} \ne 0$ & 2.37 & 2.38 & 2.37  \\
   & $B_{\mathrm{ex}} = 0$   & 2.37 & 2.38 & 2.37  \\
%%%
\end{tabular}
\end{ruledtabular}
\end{table}

% Wide table planned for two columns
\begin{table}
\caption{Orbital magnetic moments \mo\ (in \mB) inside atomic
  spheres around Co adatoms on noble metals, with magnetization
  perpendicular to the surface.  Otherwise, as for table
  \protect\ref{tab-spn-ada}. }
\label{tab-orb-ada}
%%%%%%%\begin{indented} 
%%%\lineup 
\begin{ruledtabular}
\begin{tabular}{llccc}
   &  & 
    \multicolumn{1}{c}{$\xi_{\mathrm{Co}} \ne 0$} & 
    \multicolumn{1}{c}{$\xi_{\mathrm{Co}} \ne 0$} & 
    \multicolumn{1}{c}{$\xi_{\mathrm{Co}} = 0$} 
  \\
   \multicolumn{2}{l}{substrate} & 
    \multicolumn{1}{c}{$\xi_{\mathrm{sub}} \ne 0$} & 
    \multicolumn{1}{c}{$\xi_{\mathrm{sub}} = 0$} & 
    \multicolumn{1}{c}{$\xi_{\mathrm{sub}} \ne 0$}  \\
%%%\hline  \\  [-3.5ex]
\hline
%%%
Cu & $B_{\mathrm{ex}} \ne 0$ &  1.092  & 1.095 & -0.010 \\
   & $B_{\mathrm{ex}} = 0$   &  1.096  & 1.099 & -0.010 \\
  [0.7ex]
%%%                                   
Ag & $B_{\mathrm{ex}} \ne 0$ &  1.619  & 1.611 & -0.017 \\
   & $B_{\mathrm{ex}} = 0$   &  1.619 & 1.612 & -0.017 \\
  [0.7ex]
%%%                                   
Au & $B_{\mathrm{ex}} \ne 0$ &  1.388  & 1.392 & -0.048 \\
   & $B_{\mathrm{ex}} = 0$   &  1.389  & 1.393 & -0.049 \\
  [0.7ex]
%%%                                   
Pd & $B_{\mathrm{ex}} \ne 0$ &  0.788  & 0.794 & -0.018 \\
   & $B_{\mathrm{ex}} = 0$   &  0.795  & 0.803 & -0.022 \\
  [0.7ex]
%%%                                   
Pt & $B_{\mathrm{ex}} \ne 0$ &  0.748  & 0.759 & -0.054 \\
   & $B_{\mathrm{ex}} = 0$   &  0.750  & 0.762 & -0.056 \\
%%%
\end{tabular}
\end{ruledtabular}
\end{table}

We start by investigating how the spin magnetic moment \msp\ and
orbital magnetic moment \mo\ are affected by manipulations with the
SOC parameter for Co atoms and for the substrate and with \Bex\ field
for the host.  In particular, we calculated \msp\ and \mo\ (i)~if the
SOC is fully accounted for, (ii)~if the SOC is included only on Co
atoms, and (iii)~if the SOC is included only on the substrate atoms.
For all three cases, we further distinguish the situations when the
effective exchange field in the substrate is retained
(\mm{B_{\mathrm{ex}}\ne0}) and when it is suppressed
(\mm{B_{\mathrm{ex}}=0}).  Moreover, we performed also calculations
for free-standing Co monolayers with the same geometry as the (111)
layer of the respective substrate.  The results for the adatoms are
shown in table~\ref{tab-spn-ada} and \ref{tab-orb-ada}, the results
for the monolayers are shown in table~\ref{tab-spn-lay} and
\ref{tab-orb-lay}.  For the purpose of investigating magnetic moments,
we restrict ourselves only to the case when the magnetization is
oriented perpendicular to the surface.

% Wide table planned for two columns
\begin{table}
\caption{Spin  magnetic moments \msp\ inside atomic spheres
  around Co atoms in a monolayer on a noble metal.  As for 
  table~\protect\ref{tab-spn-ada}, 
just with additional results for free-standing Co monolayers.  }
\label{tab-spn-lay}
\begin{ruledtabular}
\begin{tabular}{llcccc}
   &  & 
%%%%
    \multicolumn{1}{c}{$\xi_{\mathrm{Co}} \ne 0$} & 
    \multicolumn{1}{c}{$\xi_{\mathrm{Co}} \ne 0$} & 
    \multicolumn{1}{c}{free}  & 
    \multicolumn{1}{c}{$\xi_{\mathrm{Co}} = 0$}  \\
%%%%
 \multicolumn{2}{l}{substrate} & 
%%%%
    \multicolumn{1}{c}{$\xi_{\mathrm{sub}} \ne 0$} & 
    \multicolumn{1}{c}{$\xi_{\mathrm{sub}} = 0$} & 
    \multicolumn{1}{c}{standing} & 
    \multicolumn{1}{c}{$\xi_{\mathrm{sub}} \ne 0$} \\
\hline 
%%%
Cu & $B_{\mathrm{ex}} \ne 0$ & 1.68 & 1.67 &      & 1.67 \\
   & $B_{\mathrm{ex}} = 0$   & 1.68 & 1.67 & 1.91 & 1.67 \\
  [0.7ex]
%%%                                                
Ag & $B_{\mathrm{ex}} \ne 0$ & 1.92 & 1.92 &  & 1.92 \\
   & $B_{\mathrm{ex}} = 0$   & 1.92 & 1.92 & 2.06 & 1.92  \\
  [0.7ex]
%%%                                                
Au & $B_{\mathrm{ex}} \ne 0$ & 1.93 & 1.93 &  & 1.93  \\
   & $B_{\mathrm{ex}} = 0$   & 1.93 & 1.93 & 2.06 & 1.93 \\
  [0.7ex]
%%%                                                
Pd & $B_{\mathrm{ex}} \ne 0$ & 1.99 & 1.99 &  & 1.99  \\
   & $B_{\mathrm{ex}} = 0$   & 2.00 & 2.00 & 2.02 & 2.00 \\
  [0.7ex]
%%%                                                
Pt & $B_{\mathrm{ex}} \ne 0$ & 1.97 & 1.98 &  & 1.97  \\
   & $B_{\mathrm{ex}} = 0$   & 1.98 & 1.99 & 2.02 & 1.97 \\
%%%
\end{tabular}
\end{ruledtabular}
\end{table}

% Wide table planned for two columns
\begin{table}
\caption{Orbital magnetic moments \mo\ inside atomic spheres
  around Co atoms in a monolayer on a noble metal.  As for 
  table~\protect\ref{tab-orb-ada}, 
just with additional results for free-standing Co monolayers.  }
\label{tab-orb-lay}
\begin{ruledtabular}
\begin{tabular}{llcccc}
   &  & 
%%%%
    \multicolumn{1}{c}{$\xi_{\mathrm{Co}} \ne 0$} & 
    \multicolumn{1}{c}{$\xi_{\mathrm{Co}} \ne 0$} & 
    \multicolumn{1}{c}{free}  & 
    \multicolumn{1}{c}{$\xi_{\mathrm{Co}} = 0$}  \\
%%%%
 \multicolumn{2}{l}{substrate} & 
%%%%
    \multicolumn{1}{c}{$\xi_{\mathrm{sub}} \ne 0$} & 
    \multicolumn{1}{c}{$\xi_{\mathrm{sub}} = 0$} & 
    \multicolumn{1}{c}{standing} & 
    \multicolumn{1}{c}{$\xi_{\mathrm{sub}} \ne 0$}   \\
\hline 
%%%
Cu & $B_{\mathrm{ex}} \ne 0$  & 0.097 & 0.101 &   & -0.004 \\
   & $B_{\mathrm{ex}} = 0$   & 0.097 & 0.101 & 0.108 & -0.004 \\
  [0.7ex]
%%%                                                
Ag & $B_{\mathrm{ex}} \ne 0$   & 0.190 & 0.193 &  & -0.004 \\
   & $B_{\mathrm{ex}} = 0$    & 0.192 & 0.193 & 0.204 & -0.004 \\
  [0.7ex]
%%%                                                
Au & $B_{\mathrm{ex}} \ne 0$ & 0.171 & 0.191 &  & -0.021 \\
   & $B_{\mathrm{ex}} = 0$    & 0.171 & 0.191 & 0.202 & -0.021 \\
  [0.7ex]
%%%                                                
Pd & $B_{\mathrm{ex}} \ne 0$  & 0.155 & 0.167 &  & -0.010 \\
   & $B_{\mathrm{ex}} = 0$    & 0.155 & 0.165 & 0.160 & -0.010 \\
  [0.7ex]
%%%                                                
Pt & $B_{\mathrm{ex}} \ne 0$   & 0.141 & 0.162 &  & -0.019 \\
   & $B_{\mathrm{ex}} = 0$    & 0.141 & 0.162 & 0.164 & -0.018 \\
%%%
\end{tabular}
\end{ruledtabular}
\end{table}

It follows from our results that the magnetic moments are in many
respects quite inert 
 with respect to a manipulation of SOC and \Bex\ of the
  substrate.
By suppressing the SOC and/or the exchange field B$_{ex}$ in the
substrate, \msp\ in the Co atomic spheres changes by no more than
0.5\%.  Likewise, suppressing B$_{ex}$ in the substrate changes
\mo\ in Co atomic spheres by no more than 1\%.  This applies both for
the adatoms and for the monolayers.  If the SOC of the substrate is
suppressed, \mo\ in Co atomic spheres always increases (with the
exception of a Co adatom on Ag); this increase is at most 15\%.
Interestingly, if the SOC is included only for the substrate atoms,
the \mo\ at Co atoms arising via hybridization with substrate
SOC-split states is always negative.

The variation in \msp\ for free-standing Co monolayers reflects the
variation in the lattice constants of the substrates to which the
geometries are adjusted (cf.\ table~\ref{tab-subs}).  The largest
decrease of \msp\ due to the hybridization between Co and noble metal
states is for the Cu substrate (about 10\%), the smallest decrease is
for the Pd and Pt substrates (about 2\%).  The change in \mo\ induced
by the Co-substrate hybridization is similar as for \msp\ (less than
10\%).  This reflects the fact that we are dealing with perpendicular
magnetization here, meaning that even for supported monolayers the
quenching of \mo\ is mainly due to the hybridization with states
associated with Co atoms.

%%%------------------%%%------------------%%%---------------%%%

\subsection{Influence of SOC and \Bex\ on the MAE}

\label{sec-role-mae}

% Table planned for a single column
\begin{table}
\caption{$E_{\mathrm{MAE}} = E^{(x)}-E^{(z)}$ (in meV) for Co adatoms
  on noble metals calculated for different ways of inluding the SOC
  and \Bex.  Similarly to table~\protect\ref{tab-spn-ada}, the results
  in first colum were obtained with the SOC included at the Co atoms
  as well as at the substrate, results in the the second column are
  for the SOC included only at Co atoms, and the results in the third
  column are for the SOC included only at the substrate.  For each
  substrate, the first line shows results obtained when no
  restrictions were laid on the exchange field \Bex\ while the second
  line shows results obtained when \Bex\ was suppressed in the
  substrate. }
\label{tab-mae-ada}
\begin{ruledtabular}
\begin{tabular}{@{}llrrr}
   &  & 
    \multicolumn{1}{c}{$\xi_{\mathrm{Co}} \ne 0$} & 
    \multicolumn{1}{c}{$\xi_{\mathrm{Co}} \ne 0$} & 
    \multicolumn{1}{c}{$\xi_{\mathrm{Co}} = 0$}  
 \\
 \multicolumn{2}{l}{substrate} & 
    \multicolumn{1}{c}{$\xi_{\mathrm{sub}} \ne 0$} & 
    \multicolumn{1}{c}{$\xi_{\mathrm{sub}} = 0$} & 
    \multicolumn{1}{c}{$\xi_{\mathrm{sub}} \ne 0$}  
 \\
\hline
%%%
Cu & $B_{\mathrm{ex}} \ne 0$ & 13.17 & 12.43 & -0.03  \\
   & $B_{\mathrm{ex}} = 0$   & 13.23 & 12.50 & -0.03  \\
  [0.7ex]
%%%                                                   
Ag & $B_{\mathrm{ex}} \ne 0$ & 15.87 & 14.54 & -0.05  \\
   & $B_{\mathrm{ex}} = 0$   & 15.88 & 14.54 & -0.05  \\
  [0.7ex]
%%%                                                   
Au & $B_{\mathrm{ex}} \ne 0$ & 14.72 & 10.13 & -0.45  \\
   & $B_{\mathrm{ex}} = 0$   & 14.73 & 10.14 & -0.45  \\
  [0.7ex]
%%%                                                   
Pd & $B_{\mathrm{ex}} \ne 0$ &  6.58 &  4.65 &  0.01  \\
   & $B_{\mathrm{ex}} = 0$   &  6.49 &  4.56 &  0.03  \\
  [0.7ex]
%%%                                                   
Pt & $B_{\mathrm{ex}} \ne 0$ &  8.72 &  5.74 &  1.19  \\
   & $B_{\mathrm{ex}} = 0$   &  8.70 &  5.69 &  1.19  \\
%%%
\end{tabular}
\end{ruledtabular}
\end{table}

Our main focus is the MAE, which we calculated for the same manifold
of SOC and \Bex\ options for which we calculated \msp\ and
\mo\ above.  The results are presented in
table~\ref{tab-mae-ada} for the adatoms and in table~\ref{tab-mae-lay}
for the monolayers.  Let us recall that results obtained with full SOC
are shown in the first column of numbers, results obtained if the SOC
at the substrate is suppressed are in the second column.  So the role
of the substrate can be assessed by comparing the numbers in the first
and in the second column and, to account also for the influence of the
exchange field, the numbers in the first column should be taken for 
\mm{B_{\mathrm{ex}}\ne0} while the numbers in the second column should
be taken for \mm{B_{\mathrm{ex}}=0}.  
Table~\ref{tab-mae-lay} shows also the MAE for a free-standing Co
monolayer with the geometry of the respective substrate.  By comparing
this value with the number to the left of it we get an idea how the
MAE is influenced solely by hybridization of Co states with noble
metal states, without any contribution from the substrate SOC or 
\Bex. 
The numbers in the last column
are less important but still interesting: they represent
something that could be viewed as a ``bare''
influence of the substrate, if there is no SOC at the Co atoms.

% Table planned for 1 column
\begin{table}
\caption{$E_{\mathrm{MAE}} = E^{(x)}-E^{(z)}$ (in meV) for Co monolayers 
  on noble metals calculated for different ways of inluding the SOC
  and \Bex. 
  As for table~\protect\ref{tab-mae-ada}, 
just with additional results for free-standing Co monolayers. }
\label{tab-mae-lay}
\begin{ruledtabular}
\begin{tabular}{@{}lllcccc}
   &  & 
    \multicolumn{1}{c}{$\xi_{\mathrm{Co}} \ne 0$} & 
    \multicolumn{1}{c}{$\xi_{\mathrm{Co}} \ne 0$} & 
    \multicolumn{1}{c}{free} & 
    \multicolumn{1}{c}{$\xi_{\mathrm{Co}} = 0$}  
 \\
 \multicolumn{2}{l}{substrate} & 
    \multicolumn{1}{c}{$\xi_{\mathrm{sub}} \ne 0$} & 
    \multicolumn{1}{c}{$\xi_{\mathrm{sub}} = 0$} & 
    \multicolumn{1}{c}{standing} & 
    \multicolumn{1}{c}{$\xi_{\mathrm{sub}} \ne 0$}  
\\
\hline 
%%%
Cu & $B_{\mathrm{ex}} \ne 0$ & -0.68 & -0.83 &  & -0.02  \\
   & $B_{\mathrm{ex}} = 0$   & -0.69 & -0.83 & -1.20 & -0.02  \\
  [0.7ex]
%%%                                                   
Ag & $B_{\mathrm{ex}} \ne 0$ & -1.59 & -1.90 &  & -0.01  \\
   & $B_{\mathrm{ex}} = 0$   & -1.62 & -1.90 & -2.60 & -0.01  \\
  [0.7ex]
%%%                                                   
Au & $B_{\mathrm{ex}} \ne 0$ & -0.62 & -1.51 &  & -0.25  \\
   & $B_{\mathrm{ex}} = 0$   & -0.63 & -1.51 & -2.56 & -0.26  \\
  [0.7ex]
%%%                                                   
Pd & $B_{\mathrm{ex}} \ne 0$ &  0.20 & -0.27 &  & -0.10  \\
   & $B_{\mathrm{ex}} = 0$   &  0.15 & -0.34 & -1.83 & -0.18  \\
  [0.7ex]
%%%                                                   
Pt & $B_{\mathrm{ex}} \ne 0$ &  0.08 & -0.21 &  & -1.03  \\
   & $B_{\mathrm{ex}} = 0$   & -0.24 & -0.26 & -1.90 & -1.43  \\
%%%
\end{tabular}
\end{ruledtabular}
\end{table}

By inspecting these tables, one can recognize several general trends.
First, one can see that for the adatoms the contribution of the
substrate SOC and \Bex\  is
relatively small, while for the monolayers this contribution
can be sometimes truly substantial.  To be more specific, the situation
for the adatoms is such that the magnetic easy axis is always
out-of-plane, no matter whether the substrate SOC and \Bex\ is
included or not; the effect of switching on the substrate is just that
the numerical values for the \mae\ increase  (by 5\% for the Cu
substrate, by 35\% for the Pt substrate).  For the monolayers, on the
other hand, including the substrate SOC and/or \Bex\ may reorient the
magnetic easy axis: it is in-plane if the substrate contribution is
suppressed but it is rotated to the out-of-plane direction if the
substrate SOC and \Bex\ is included for the Pd and Pt substrates.
(For Cu, Ag, and Au substrates the easy magnetic axis of a Co
monolayer remains in-plane if the substrate SOC and \Bex\ are switched
on but the absolute value of the \mae\ decreases.)

It follows also from Tabs.~\ref{tab-mae-ada}--\ref{tab-mae-lay} that
the contribution due to the substrate SOC is practically always more
important than the contribution due to the substrate exchange field
$B_{\mathrm{ex}}$.  In particular, suppressing substrate
$B_{\mathrm{ex}}$ has practically no effect for the adatoms.  For the
monolayers, \Bex\ has got a negligible influence in case of Cu, Ag,
and Au substrates, a significant influence in case of the Pd substrate
and a crucial influence in case of the the Pt substrate (indeed, it is
the substrate exchange field that switches the magnetic easy axis from
in-plane to out-of-plane).

Even though our focus is on SOC and \Bex, it is instructive to have a
look at the changes in the MAE caused by depositing a free-standing Co
monolayer on a substrate with $\xi_{\mathrm{sub}}$=0 and \Bex=0.  This
could be viewed as the pure effect of Co-substrate hybridization.
The strength of this effect appears to be significantly larger for the
Pd and Pt substrates than for the Cu, Ag, and Au substrates.  This
seems to reflect the fact that the overlap between Co and noble metal
valence bands is larger for the Co/Pd and Co/Pt interfaces on the one
hand than for the Co/Cu, Co/Ag, and Co/Au interfaces on the other
hand \cite{WWF93a,DKS+94,SBM+07}. Different roles of hybridization in
this respect were discussed in detail by Wang \ea \cite{WWF93a} and by
Daalderop \ea \cite{DKS+94}.

Let us note finally that 
by comparing the \mae\ values in the first, the second and the last
column in Tabs.~\ref{tab-mae-ada}--\ref{tab-mae-lay}, one sees
immediately that the effect of the SOC at different sites is not
additive: the true MAE is clearly not a sum of the MAE obtained if
the SOC is included only at the Co atoms with the MAE obtained if the
SOC is included only at the substrate --- not even in case of
substrates with weak or moderate SOC.

%%%------------------%%%------------------%%%---------------%%%

\subsection{Comparing SOC and \Bex\ manipulation with
  decomposition of \mae\ by means of the torque contributions}

As it was mentioned in the introduction, the torque method has
often been used to resolve the MAE into
localized contributions.  It is thus instructive to compare
quantitatively the role of the substrate as deduced from the SOC and
\Bex\ manipulation and as provided by a mechanical assignment of
individual terms in the sum of the torque contributions to individual
atoms,
\begin{equation}
E_{\mathrm{MAE}} \; = \; \sum_{j}
T^{(\theta=45^{\circ})}_{j}
\label{torsum}
\;\;.
\end{equation}
In particular, using the method employed in this work, a  quantitative
measure of the role of the substrate for \mae\ can be obtained by
subtracting and dividing appropriate values in
Tabs.~\ref{tab-mae-ada}--\ref{tab-mae-lay}, which can be symbolically
written as 
\begin{equation}
w_{\mathrm{sub}}^{(\mathrm{SOC},B_{\mathrm{ex}})} \: = \:
\frac{
  E_{\mathrm{MAE}}(\{\mathrm{Co},\mathrm{sub}\}) 
  -
  E_{\mathrm{MAE}}(\{\mathrm{Co}\} ) }
  { E_{\mathrm{MAE}}(\{\mathrm{Co},\mathrm{sub}\})  }
\label{racsoc}
\end{equation}
with
\begin{eqnarray*}
& \{\mathrm{Co},\mathrm{sub}\} & \, \cong  \, 
\{ \xi_{\mathrm{Co}}\ne0, \, \xi_{\mathrm{sub}}\ne0, \, 
  B_{\mathrm{ex}}\ne0 \}
\;\;,
\\
& \{\mathrm{Co}\}  & \, \cong \, 
\{ \xi_{\mathrm{Co}}\ne0,  \, \xi_{\mathrm{sub}}=0, \, 
  B_{\mathrm{ex}}=0 \}
\;\;.
\end{eqnarray*}
To get an analogous quantity by relying on resolving the sum of the
torque contributions, one can apply a procedure that can be symbolically
denoted as 
\begin{equation}
w_{\mathrm{sub}}^{(T_{j})} \: = \:
\frac{  \sum_{\mathrm{sub}}
T^{(\theta=45^{\circ})}_{j} }
{  \sum_{\mathrm{Co},\mathrm{sub}}
T^{(\theta=45^{\circ})}_{j} }
\label{ractor}
\;\;.
\end{equation}
One should, however, keep in mind that proceeding along this second
scheme is not internally consistent as one implicitly makes an
assumption that the energy is ``spatially additive''.  Use of
equation~(\ref{racsoc}), on the other hand, is free of such issues because
now we always evaluate the energy of the whole model system --- we
only change its properties by manipulating the SOC and \Bex.

\begin{table}
\caption{Role of the substrate in generating the MAE for Co adatoms on
noble metals as assessed by SOC and $B_{\mathrm{ex}}$ manipulation and
as assessed by comparing individual terms in the torque evaluation.}
\label{tab-sub-ada}
\begin{ruledtabular}
\begin{tabular}{@{}lrr}
   &  \multicolumn{1}{c}{via SOC and \Bex} &
    \multicolumn{1}{c}{via comparing} \\
  \multicolumn{1}{c}{substrate}  &  \multicolumn{1}{c}{manipulation} &
    \multicolumn{1}{c}{$T_{i}$ terms} \\
\hline 
%%%
Cu  &  5.1\%  &  0.02\%  \\
Ag  &  8.4\%  &  0.00\%  \\
Au  & 31.1\%  &  0.01\%  \\
Pd  & 30.7\%  &  0.67\%  \\
Pt  & 34.8\%  &  0.14\%  \\
\end{tabular}
\end{ruledtabular}
\end{table}

\begin{table}
\caption{As for table~\protect\ref{tab-sub-ada}, however, for Co
  monolayers instead of Co adatoms.}
\label{tab-sub-lay}
\begin{ruledtabular}
\begin{tabular}{@{}lrr}
   &  \multicolumn{1}{c}{via SOC and \Bex} &
    \multicolumn{1}{c}{via comparing} \\
  \multicolumn{1}{c}{substrate}  &  \multicolumn{1}{c}{manipulation} &
    \multicolumn{1}{c}{$T_{i}$ terms} \\
\hline 
%%%
Cu  &   -20.4\%  &   -0.10\%  \\
Ag  &   -19.3\%  &   -0.06\%  \\
Au  & -142.5\%  &   -0.39\%  \\
Pd  & 265.5\%  &   24.7\%  \\
Pt  & 422.5\%  &  190.9\%  \\
\end{tabular}
\end{ruledtabular}
\end{table}

The relative importance of the substrate evaluated via procedures
outlined in Eqs.~(\ref{racsoc})--(\ref{ractor}) is presented in
table~\ref{tab-sub-ada} for Co adatoms and in table~\ref{tab-sub-lay}
for Co monolayers.  One sees immediately that there are clear
differences between both procedures.  The more physical approach based
on the SOC and \Bex\ manipulation reveals that the role of the
substrate is significantly larger than what would follow from the
mechanistic decomposition of the torque sum. 
In some cases this difference is striking (such as, e.g., for Co
adatoms in Au, Pd, and Pt or for a Co monolayer on Au).

%%%%%%%%%%%%%%%%%%%%%%%%%%%%%%%%%%%%%%%%%%%%%%%%%%%%%%%%%%%%%%%%%%%%%%%
%%%%%%%%%%%%%%%%%%%%%%%%%%%%%%%%%%%%%%%%%%%%%%%%%%%%%%%%%%%%%%%%%%%%%%%

\section{Discussion}

\label{sec-diss}

Our goal was to study the localization of the MAE in complex systems,
with focus on the question whether the MAE of adatoms and monolayers
adsorbed on non-magnetic supports resides mostly in the adsorbed atoms
or in the substrate.  We noted that this question in principle cannot
be answered, or at least cannot be answered in an 
unambiguous way, because the energy of an inhomogeneous system is not
an extensive quantity and thus the energy of a composed system cannot be
split into energies residing in sub-parts of the system.  
 At the same
time, the simple question ``where does the anisotropy come from''
follows naturally from the effort to understand the MAE 
in simple terms.  Therefore, it is desirable to re-formulate it 
in such a way that it does not suffer from  inconsistencies
and still reflects the intuitive question about the role of the
adsorbates and the substrates in generating the magnetocrystalline
anisotropy.  The approach we adopted, namely, comparing the MAE
calculated for the original system with the MAE calculated for a model
system where the key factors contributing to the magnetocrystalline
anisotropy (such as the SOC and \Bex) are selectively suppressed
satisfies this requirement.

% Table planned for a single column
\begin{table}
\caption{\mae\ (in meV) calculated in this work compared with
    other {\em ab initio} calculations and with experiment.  The
    systems include free-standing monolayers Co$_{\infty}$ with
    geometries of Pd(111) and Pt(111) (first two lines) and adatoms
    Co$_{1}$ and monolayers Co$_{\infty}$ supported by (111) surfaces
    of fcc substrates.  The experimental values for monolayers include
    also a dipole-dipole contribution of about
    -0.09~meV \cite{BSM+12}.}
\label{tab-compar}
\begin{ruledtabular}
\begin{tabular}{@{}lccc}
 system  &  
\multicolumn{1}{c}{this work} & 
\multicolumn{1}{c}{other theory} & 
\multicolumn{1}{c}{experiment}  \\
\hline 
%%%
Co$_{\infty}$ as Pd  &  
-1.8  &  -2$^{\rm a}$ & \mbox{---} \\
%%%
Co$_{\infty}$ as Pt  &  
-1.9  &  -1.6$^{\rm b}$  &  \mbox{---} \\
%%%
Co$_{1}$/Pd  &  6.6  &   1.9$^{\rm c}$  &  
 $\sim$3$^{\rm d}$ \\
%%%
Co$_{1}$/Pt  &  8.7  &  
\multicolumn{1}{c}{  8.1,$^{\rm e}$
                     3.1,$^{\rm f}$
                     5.0,$^{\rm g}$
                     5.9$^{\rm h}$ }
  & 9.3,$^{\rm i}$ 10$^{\rm f}$ \\
%%%
Co$_{\infty}$/Cu  &  -0.7  &  
 -0.5$^{\rm j}$ & $<0$$^{\rm k}$\\
%%%
Co$_{\infty}$/Au  &  -0.6  &  
 -0.6$^{\rm l}$ & $<0$$^{\rm m}$ \\
%%%
Co$_{\infty}$/Pt  &  0.1  &  
  0.1,$^{\rm h}$ 
  1.1$^{\rm n}$   & 
  0.15,$^{\rm i}$  
  0.12,$^{\rm o}$ 
  $>0$$^{\rm p}$  
  \\
%%%
\end{tabular}
\end{ruledtabular}
$^{\rm a}$ {Daalderop \ea \cite{DKS+94}} \\
$^{\rm b}$ {Lehnert \ea \cite{LDB+10}} \\
$^{\rm c}$ {B{\l}o\'{n}ski \ea \cite{BLD+10} (for a relaxed geometry)} \\
$^{\rm d}$ {Claude \cite{Cla05}} \\
$^{\rm e}$ {B{\l}o\'{n}ski \& Hafner \cite{BH+09} (value for
   a bulk-like geometry is shown)} \\
$^{\rm f}$ {Balashov \ea \cite{BST+09} (calculations for a 
relaxed geometry)} \\
$^{\rm g}$ {Etz \ea \cite{EZWV08}} \\
$^{\rm h}$ {Lazarovits \ea \cite{LSW03}} \\
$^{\rm i}$ {Gambardella \ea \cite{GRV+03}} \\
$^{\rm j}$ {Hammerling \ea \cite{HUZ+02}} \\
$^{\rm k}$ {Huang \ea \cite{HKM+94}} \\
$^{\rm l}$ {\'{U}jfalussy \ea \cite{USBW96}} \\
$^{\rm m}$ {Padovani \ea \cite{PSC+00}} \\
$^{\rm n}$ {Lehnert \ea \cite{LDB+10} (for a monolayer in an hcp
  position with a relaxed geometry)} \\
$^{\rm o}$ {Meier \ea \cite{MBF+06}} \\
$^{\rm p}$ {Moulas \ea \cite{MLR+08}} 
\end{table}

Before we proceed further, let us compare our results with earlier
theoretical and experimental results for the same systems we explore
here.  This is done in table~\ref{tab-compar}.  When analyzing the
theoretical results, one should have in mind that 
 comparing theoretical MAE values obtained by different studies
  is not always straightforward.  First, the MAE is sensitive to the
  adatom-substrate geometry relaxation \cite{BH+09,CFB+08}, so 
  quantitative differences between different works may be due to
  different interatomic distances used.  However, the MAE of adatoms
  and monolayers is also sensitive to the way the substrate is
  accounted for (i.e, how many layers have been used to model the
  semi-infinite half-crystal) and to whether the adatoms are allowed
  to interact with each other or not (i.e., what is the lateral size
  of the supercell which simulates the adatom) \cite{SBME10}.
Also
technical parameters such as angular momentum cutoff
$\ell_{\mathrm{max}}$ are important.  To analyze the differences between
all the various theoretical calculations would thus be quite complicated and
beyond our scope.  
 Still we should address two probably most serious
  simplifications of our treatment, which is the use of the ASA and
  the neglect of the geometry relaxation.  
For open systems such as adatoms and, to a lesser degree,
  supported monolayers the use of the ASA will certainly affect the
  values of the MAE.
However, it follows from the comparison
  between ASA and full potential calculations for identical systems
  that the effect of the ASA 
should not be crucial.  E.g., for a Co monolayer on
  Pd(100) with a bulk-like geometry, one gets MAE of -0.73~meV using
  the ASA \cite{SBE+13} and -0.75~meV using a full potential
  \cite{WCF+97}.  For a Co adatom on Pt(111) with a bulk-like
  geometry, the MAE is 8.7~meV if obtained using the ASA (this work),
  9.2~meV if obtained using a full-potential for a 4$\times$4
  supercell on a 4-layers thick slab \cite{CFB+08} and 8.1~meV if
  obtained using a full potential for a 5$\times$5 supercell on a
  5-layers thick slab \cite{BH+09}. For a Co adatom on Pt(111) in an
  adsorption hcp position with a relaxed geometry, the ASA yields MAE
  of 1.90~meV \cite{SBE+13} and a full potential yields MAE of
  0.72~meV (for a 5$\times$5 supercell on a 5-layers thick slab)
  \cite{BLD+10}.

 Our use of bulk-like geometries will probably affect the
  calculated MAE more than the ASA does.  To be quantitative, relaxing
  the geometry for a Co adatom in an fcc adsorption position on
  Pt(111) changes the MAE from 9.2~meV to 4.8~meV \cite{CFB+08}.
  Using an optimized geometry for a Co monolayer on Pd(111) instead of
  a bulk geometry changes the MAE from 0.21~meV to 0.36~meV
  \cite{SBE+13}.  Relaxing a Fe monolayer on Pt(111) changes the MAE
  from -0.66~meV to -0.47~meV \cite{THO+07}.  Similar deviations have
  to be expected for our systems.  Therefore, one has to take our
  values of the MAE with care when interpreting experiments on real
  materials.  
This is especially true for the Pt and Au substrates: atomic
  volumes of 3$d$ elements are significantly smaller than atomic
  volumes of 5$d$ elements, which will result in shorter Co--Pt and
  Co--Au distances and, consequently, smaller magnetic moments and
  smaller MAE in comparison with the values we obtained here for the
  bulk-like distances. 
One could argue that a cautious attitude should be applied
  to all LDA calculations of MAE for adatoms anyway, because of
  possible orbital polarization effects \cite{GRV+03,NCZ+01} which are
  hard to describe within conventional {\em ab initio} procedures.  The
  important thing is that our focus here is not on the particular value
  of the MAE but on the general trends over a large set systems, each
  of them being treated with the same technical parameters.  As it
  follows from table~\ref{tab-compar}, our calculations yield results
  in the same range of values as other {\em ab initio} calculations
  and also as provided by experiment, which gives us confidence
  that we can use them to draw reliable conclusions concerning the
  effect of the substrate SOC and \Bex\ when going from adatoms to
  monolayers and when going through substrates of various
  properites.

We found that, generally, the substrate is more important when dealing with
monolayers than when dealing with adatoms.  A similar observation
could be made also on the basis of several earlier studies performed
via the torque decomposion (see end of Sec.~\ref{sec-uptonow}).  On
the one hand, this is surprising, because the ratio of the number of
participating substrate atoms to the number of adsorbed atoms is much
larger for the adatoms than for the monolayers so one would expect
that as a consequence of this, the substrate should be more important
for the adatoms than for the monolayers.  On the other hand, one could
argue that the electronic structure of the (originally) non-magnetic
substrate is more altered by the presence of monolayers than by the
presence of adatoms, suggesting that the involvement of the substrate
in the magnetocrystalline anisotropy will be higher for the monolayers
than for the adatoms.  The results demonstrate that the second trend
prevails.

The exchange field \Bex\ in the substrate has practically no effect on
the MAE in case of adatoms.  This is not surprising for the Cu, Ag,
and Au substrates because they have the enhancement $S_{xc}$ factor
close to unity.  However, this holds also for the Pd and Pt
substrates which is quite surprising because these elements have quite
large $S_{xc}$ and, moreover, a 3$d$ adatom or impurity induces in
these materials an extended polarization cloud, the magnetic moment of
which may be larger than the moment of the inducing 3$d$
atom \cite{CS65,Zel93,SMM+08}.

Turning to the role of the substrate \Bex\ field for Co 
 monolayers, it is unimportant in the
case of Cu, Ag, and Au substrates.  However, it is significant in the
case of a Co monolayer on the Pd substrate and crucial in the case of a
Co monolayer on the Pt substrate (cf.\ corresponding lines in
table~\ref{tab-mae-lay} labelled by $B_{\mathrm{ex}}\ne0$ and by
$B_{\mathrm{ex}}=0$).  Interestingly, the role of the \Bex\ field is
larger for Pt than for Pd even though the $S_{xc}$ factor for Pd is
about three times larger than for Pt (table~\ref{tab-subs}).  Another
intriguing feature is that for a Co monolayer on Pt, the importance of
the substrate \Bex\ field strongly depends on whether the SOC is fully
included or whether it is included only on one type of atoms (either
on Co atoms or on Pt atoms): in the former case, the role of
\Bex\ is significantly more important than in the latter case.  We
could summarize this point by saying that the effects of B$_{ex}$ and
SOC are intertwined in this case and both factors contribute to the
MAE in an non-additive way.  
While the effect of the SOC was explored for some layered systems
already \cite{WWF93a,BRB+06,SF+11,DKS+94,SDM95,RKF+01,Cin+01,BES+05}, 
the role of the substrate \Bex\ has been investigated here for the
first time.

Let us recall again that it is in principle not possible to decompose
the MAE into a sum of site-related quantities.  This can be
illustrated also by analysis of
Tabs.~\ref{tab-mae-ada}--\ref{tab-mae-lay}, because the values in the
``$\xi_{\mathrm{Co}}\ne0$, $\xi_{\mathrm{sub}}\ne0$'' column clearly
differ from the sum of the values in the ``$\xi_{\mathrm{Co}}\ne0$,
$\xi_{\mathrm{sub}}=0$'' and in the ``$\xi_{\mathrm{Co}}=0$,
$\xi_{\mathrm{sub}}\ne0$'' columns.  The fact that one cannot decompose
the MAE into a sum of contributions corresponding to situations 
where the SOC is included only on one atomic type at a time was
pointed out already in some earlier works, e.g., by Wang
\ea \cite{WWF93a} for a Pd/Co/Pd sandwich or by Subkow and
F\"{a}hnle \cite{SF+11} for a Fe-Au interface.

Another interesting point in this respect is that a stronger SOC for a
substrate does not necessarily mean that it has got a higher relative
importance concerning the MAE.  In particular, the SOC for Ag is
about twice as strong as for Cu and yet the relative role of these
substrates for the MAE of a Co {\em monolayer} supported by them is
the same --- about 20\% (table~\ref{tab-sub-lay}).  The \Bex\ field
does not interfere here because its role is negligible both for Cu and
for Ag (table~\ref{tab-mae-lay}).  
The relatively small role of the SOC for the Ag substrate reminds a
similar situation for Co/Ag multilayers: Daalderop \ea \cite{DKS+94}
found that even though the SOC strength is similar for Pd and Ag, its
role is more significantly important for the Co$_{1}$Pd$_{2}$
multilayers than for the Co$_{1}$Ag$_{2}$ multilayers.
In other comparisons, however, it appears that stronger SOC indeed
implies a bigger role of the substrate for the MAE (cf.\ Cu, Ag, and
Au substrates for a Co adatom, table~\ref{tab-sub-ada}).  So it seems
that there is no unique pattern in this respect.

It follows from our results that if one analyzes the effects of
site-related SOC and \Bex\ for the MAE, the role of the substrate is
much more important than what one gets from comparing individual
site-related terms in the torque evaluation.  Especially this is true
for Co adatoms on Au, Pd, and Pt and for a Co monolayer on Au, where
the differences are two orders of magnitude.  So while evaluating the
torque is a convenient way to calculate the MAE of a system, it
should not be used for assessing the roles of various constituents for
the magnetocrystalline anisotropy of a compound or a nanostructure.

Although we have not explicitly tested for our systems the
  decomposition scheme based on site-projected densities of states, we
  expect that the outcome concerning the role of the substrate would
  be similar as with the torque formula, among others because both
  approaches are based on the magnetic force theorem.  This view is
  based also on the analysis of the results of works which employed
  this scheme: By decomposing the MAE for surfaces and multilayers
  into layer-resolved contributions via site-projected DOS it was
  found that the main contribution comes from surfaces and interfaces,
  with only a small part coming from non-magnetic substrates or
  spacers \cite{USW96,USBW96,CE+97,HUZ+02}. Such an outcome clearly
  differs from the picture obtained via site-related SOC and
  \Bex\ analysis in the present work.

All the substrate materials we investigated had the tendency to orient
the magnetic easy axis in the out-of-plane direction: by switching on
the SOC and \Bex\ in the substrate, either the out-of-plane
orientation of the magnetic easy axis was reinforced (in the case of
adatoms, see table~\ref{tab-mae-ada}), or the preference of the
magnetic easy axis for the in-plane orientation got weaker (in case of
monolayers on Cu, Ag, and Au, see table~\ref{tab-mae-lay}), or the
magnetic easy axis was re-oriented from the in-plane direction to the
out-of-plane direction (in case of monolayers on Pd and Pt).  It would
be interesting to check for other adsorbates how general this tendency
is.  Finally, it should be noted that the same approach we used here
could be applied also to layered systems such as CoPt or FePd to
assess the role of the non-magnetic element in these systems.

%%%%%%%%%%%%%%%%%%%%%%%%%%%%%%%%%%%%%%%%%%%%%%%%%%%%%%%%%%%%%%%%%%%%%%%

\section{Conclusions}

The role of the substrate for generating the magnetocrystalline
anisotropy of supported nanostructures can be assessed by comparing
the MAE calculated for the real system with the MAE calculated for a
model system where the spin orbit coupling and the effective exchange
field $B_{\mathrm{ex}}$ is suppressed at the substrate atoms.  For Co
adatoms on noble metals (Cu, Ag, Au, Pd, Pt), the 
 contribution of the substrate SOC and \Bex\ 
to the MAE is relatively small while for Co monolayers it
can be substantial.  For all five substrates we explored, we found
that their contribution to the MAE is out-of-plane.

The role of the substrate SOC is more important than the role of the
substrate exchange field $B_{\mathrm{ex}}$.  For Co adatoms on Cu, Ag,
Au, Pd, or Pt, the substrate \Bex\ field has practically no effect on
the MAE.  For Co monolayers, the substrate \Bex\ field is unimportant
for substrates which are hard to polarize (Cu, Ag, Au) but it is
significant for highly polarizable substrates (Pd, Pt).  Generally,
the effects of the SOC and of the $B_{\mathrm{ex}}$ field effect are
non-additive.  The same is true for the effect of SOC if it is
selectively switched on either only for the adsorbed atoms or only for
the substrate atoms.

%%%%%%%%%%%%%%%%%%%%%%%%%%%%%%%%%%%%%%%%%%%%%%%%%%%%%%%%%%%%%%%

\begin{acknowledgments}
This work was supported by the Grant Agency of the Czech Republic
within the project 108/11/0853, by the Bundesministerium f\"{u}r
Bildung und Forschung (BMBF) project 05K13WMA, and by the Deutsche
Forschungsgemeinschaft (DFG) within the SFB~689.
\end{acknowledgments}

%%%%%%%%%%%%%%%%%%%%%%%%%%%%%%%%%%%%%%%%%%%%%%%%%%%%%%%%%%%%%%%
\appendix

%%%%%%%%%%%%%%%%%%%%%%%%%%%%%%%%%%%%%%%%%%%%%%%%%%%%%%%%%%%%%%%

\section{Convergence of the MAE with respect to the
  $\mathbf{k}$ space integration grid}

\label{sec-kgrid}

% Narrow table planned for a single column
\begin{table}
\caption{Dependence of the MAE for a Co monolayer on Cu(111) and on
  Pt(111) on the number $N_{k}$ of integration points in the surface
  Brillouin zone. The MAE was calculated both by evaluating the torque
  according to equation (\protect\ref{MAEvalue}) (columns labelled by 
``$\partial E(\theta)/\partial \theta$'') and directly by
  subtracting the total energies for two orientations of the
  magnetization (columns labelled by ``$\Delta
  E_{\mathrm{tot}}$''). The MAE is in meV's.} 
\label{tab-kpoints-mono}
\begin{ruledtabular}
\begin{tabular}{rllll}
 & \multicolumn{1}{c}{Cu} & \multicolumn{1}{c}{Cu} & 
   \multicolumn{1}{c}{Pt} & \multicolumn{1}{c}{Pt}  \\
 \multicolumn{1}{c}{$N_{k}$}  &  
 \multicolumn{1}{c}{
      $\frac{\partial E(\theta)}{\partial \theta}$}
 & \multicolumn{1}{c}{$\Delta E_{\mathrm{tot}}$}
 &  \multicolumn{1}{c}{
    $\frac{\partial E(\theta)}{\partial \theta}$}
 & \multicolumn{1}{c}{$\Delta E_{\mathrm{tot}}$}
  \\
%%%\hline  \\  [-3.5ex]
\hline
%%%
  3600 &  -0.684 &   -0.720 &   0.115  & 0.012  \\
  6400 &  -0.650 &   -0.696 &   0.198  & 0.202  \\
 10000 &  -0.685 &   -0.714 &   0.080  & 0.100  \\
 22500 &  -0.688 &   -0.732 &   0.128  & 0.112  \\
 40000 &  -0.687 &   -0.726 &   0.113  & 0.120  \\
%%%
\end{tabular}
\end{ruledtabular}
\end{table}

% Narrow table planned for a single column
\begin{table}
\caption{Dependence of the MAE for a Co adatom on Cu(111) and on
  Pt(111) on the number $N_{k}$ of integration points in the surface
  Brillouin zone used when calculating the electronic structure and
  the Green's function of the host. The MAE is in meV's. }
\label{tab-kpoints-adatom}
\begin{ruledtabular}
\begin{tabular}{rll}
 & \multicolumn{1}{c}{Cu} &  
   \multicolumn{1}{c}{Pt}  \\
 \multicolumn{1}{c}{$N_{k}$}  &  
 \multicolumn{1}{c}{
      $\frac{\partial E(\theta)}{\partial \theta}$}
 &  \multicolumn{1}{c}{
    $\frac{\partial E(\theta)}{\partial \theta}$}
  \\
%%%\hline  \\  [-3.5ex]
\hline
%%%
  3600  &  12.046  &  8.474   \\
  6400  &  12.395  &  8.692   \\
 10000  &  13.170  &  8.719   \\
 22500  &  12.909  &  8.689   \\
 40000  &  13.181  &  8.763   \\
%%%
\end{tabular}
\end{ruledtabular}
\end{table}
 
% Figure planned for 2 columns
\begin{figure*}
\includegraphics[viewport=0 0 484 343,width=160mm]{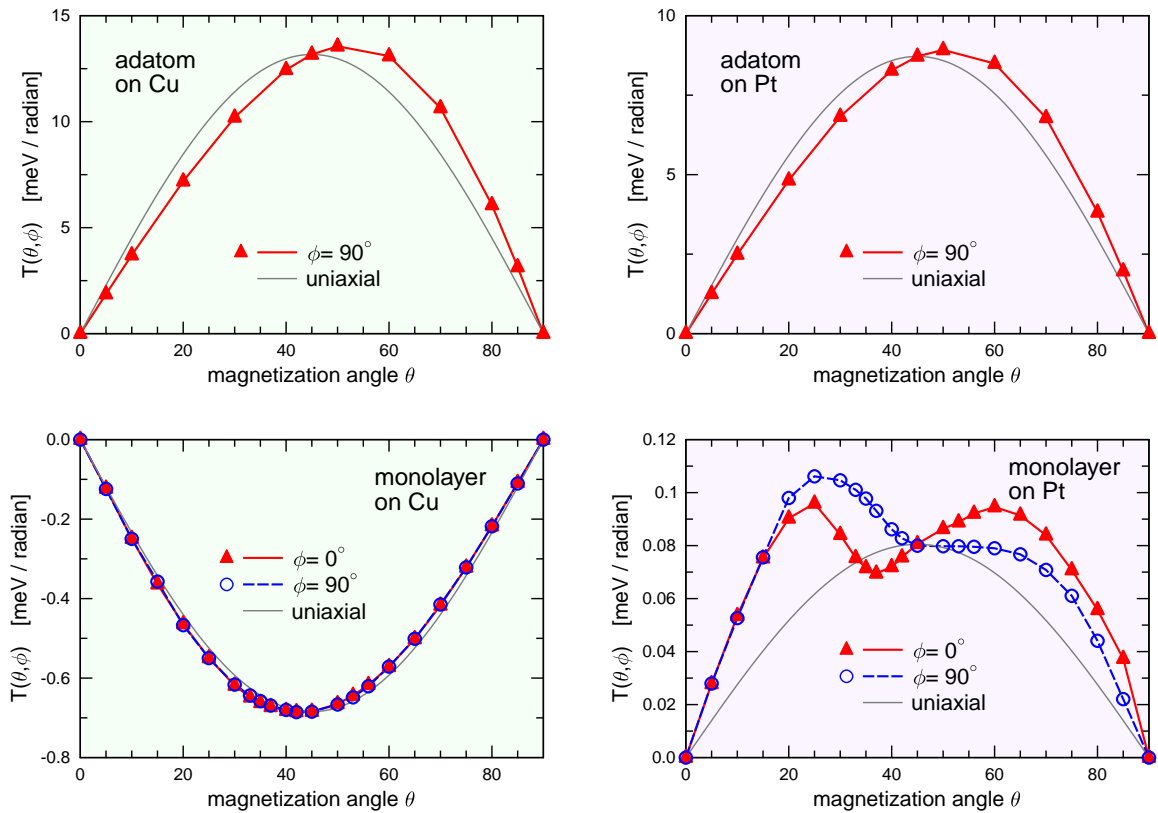}%
\caption{(Color online) Torque $T(\theta,\phi)$ as a function of
  $\theta$ for Co adatoms and monolayers on Cu(111) and Pt(111).
  Markers represent calculated points, thick lines are guides for an
  eye, thin lines are uniaxial contributions [proportional to
    $\sin(2\theta)$] which would yield the same MAE as provided by
  equation (\protect\ref{MAEvalue}). }
\label{fig-torque}
\end{figure*}

 The integration within the 2-dimensional $\mathbf{k}$ space was
  done on a regular grid.  We used 10000 points in the full surface
  Brillouin zone to determine the self-consistent potentials as well
  as to evaluate the MAE via equation (\ref{MAEvalue}).  According to 
  our experience, this grid density is in the regime where the results
  are already converged enough to provide the MAE with accuracy of
  about 5\% (for very small absolute values of the MAE, the relative
  accuracy may be somewhat  worse).  We demonstrate this in
  tables~\ref{tab-kpoints-mono}--\ref{tab-kpoints-adatom} where we
  explore the $\mathbf{k}$-grid dependence of the MAE for two extreme
  cases of the substrate material, namely, for low-SOC low-polarizable
  Cu and high-SOC highly-polarizable Pt.  Apart from results obtained
  by evaluating the torque according to equation (\ref{MAEvalue}), we
  present for the monolayers also results obtained by a direct
  subtraction of the total energies for two orientations of the
  magnetization.  This demonstrates that relying on the magnetic force
  theorem together with equation (\ref{MAEvalue}) is justified for our
  purpose.

%%%%%%%%%%%%%%%%%%%%%%%%%%%%%%%%%%%%%%%%%%%%%%%%%%%%%%%%%%%%%%%

\section{Higher-order contributions to the MAE}

\label{sec-torque}

 When using equation (\ref{MAEvalue}), we assume a simple
  uniaxial dependence of the torque on the polar angle $\theta$, i.e.,
  we neglect higher-order contributions.  These can be accounted for
  if we evaluate the torque $T(\theta,\phi)$ for the whole $\theta$
  range.  The anisotropy then can be evaluated by performing the
  integral 
\begin{equation}
E_{\mathrm{MAE}} \, = \, \int_{0}^{\pi/2} \dstd \theta \: T(\theta,\phi)
\;\;.  
\label{torint}
\end{equation}
 To verify that it is sufficient for our purpose to restrict
  ourselves to uniaxial contributions, we present here the full
  $\theta$ scan of the torque $T(\theta,\phi)$ for Co adatoms and
  monolayers on Cu(111) and on Pt(111).  For the monolayers we probe
  the azimuthal dependence as well, i.e., we perform the $\theta$
  scans for the magnetization direction confined either to the $xz$
  plane ($\phi=0^{\circ}$, horizontal direction in
  figure~\ref{fig-geom}) or to the $yz$ plane ($\phi=90^{\circ}$,
  vertical direction in figure~\ref{fig-geom}).  The azimuthal
  dependence of the MAE for the adatoms on fcc (111) surfaces is
  expected to be very weak (cf.\ data for a Co adatom on Pd(111)
  \cite{SBE+13}) so it is not explored here --- just data for
  $\phi=90^{\circ}$ are presented in this case.

% Wide table planned for two columns
\begin{table*}
\caption{MAE for Co adatoms and monolayers on Cu(111) and Pt(111)
  evaluated just using the torque at $\theta$=45$^{\circ}$ according
  to equation (\protect\ref{MAEvalue}) and using the torque over the
  whole $\theta$ range according to equation (\protect\ref{torint}).
  The MAE is in meV's. }
\label{tab-uni}
\begin{ruledtabular}
\begin{tabular}{lcccccc}
 & \multicolumn{1}{c}{adatom} & \multicolumn{1}{c}{monolayer} & 
   \multicolumn{1}{c}{monolayer} & \multicolumn{1}{c}{adatom} & 
   \multicolumn{1}{c}{monolayer} & \multicolumn{1}{c}{monolayer} \\
 & \multicolumn{1}{c}{on Cu}  & \multicolumn{1}{c}{on Cu}  & 
   \multicolumn{1}{c}{on Cu}  & \multicolumn{1}{c}{on Pt}  & 
   \multicolumn{1}{c}{on Pt}  & \multicolumn{1}{c}{on Pt}  \\
 & \multicolumn{1}{c}{$\phi=90^{\circ}$}  &  
   \multicolumn{1}{c}{$\phi=0^{\circ}$}  &  
   \multicolumn{1}{c}{$\phi=90^{\circ}$} & 
   \multicolumn{1}{c}{$\phi=90^{\circ}$}  &  
   \multicolumn{1}{c}{$\phi=0^{\circ}$}  &  
   \multicolumn{1}{c}{$\phi=90^{\circ}$} \\
%%%\hline  \\  [-3.5ex]
\hline
%%%
via eq.~(\ref{MAEvalue})   &  
 13.170 & -0.684 & -0.685 &  8.719 & 0.081 & 0.080 \\
via eq.~(\ref{torint})   &  
 13.570 &  -0.685 & -0.684 & 8.897 & 0.111 & 0.109 \\
%%%
\end{tabular}
\end{ruledtabular}
\end{table*}

 The dependence of the torque on $\theta$ is shown in
  figure~\ref{fig-torque}.  The comparison of the MAE obtained
  via equation (\ref{MAEvalue}) and via equation (\ref{torint}) is
  shown in table~\ref{tab-uni}.  One can see from
  figure~\ref{fig-torque} that there are clear deviations from the
  simple uniaxial behaviour of $T(\theta,\phi)$ both for adatoms and
  for monolayers.  These deviations are especially evident for the Co
  monolayer on Pt(111), which is probably connected with the large
  role of the substrate in this case (see section~\ref{sec-role-mae}).
  However, the data presented in table~\ref{tab-uni} demonstrate that
  as long as we are concerned with integral quantities such as is the
  MAE defined in equation~(\ref{maedef}), it is sufficient in our case
  to deal just with the uniaxial contributions, i.e., to rely on
  equation (\ref{MAEvalue}).

%%%%%%%%%%%%%%%%%%%%%%%%%%%%%%%%%%%%%%%%%%%%%%%%%%%%%%%%%%%%%%%
%%      References

%%%\section*{References}

% Produces the bibliography via BibTeX.
% File *.bbl has to be inserted manually 

\bibliography{liter_local-MAE}

\begin{thebibliography}{10}%
\makeatletter
\providecommand \@ifxundefined [1]{%
 \ifx #1\undefined \expandafter \@firstoftwo
 \else \expandafter \@secondoftwo
\fi
}%
\providecommand \@ifnum [1]{%
 \ifnum #1\expandafter \@firstoftwo
 \else \expandafter \@secondoftwo
\fi
}%
\providecommand \enquote [1]{``#1''}%
\providecommand \bibnamefont  [1]{#1}%
\providecommand \bibfnamefont [1]{#1}%
\providecommand \citenamefont [1]{#1}%
\providecommand\href[0]{\@sanitize\@href}%
\providecommand\@href[1]{\endgroup\@@startlink{#1}\endgroup\@@href}%
\providecommand\@@href[1]{#1\@@endlink}%
\providecommand \@sanitize [0]{\begingroup\catcode`\&12\catcode`\#12\relax}%
\@ifxundefined \pdfoutput {\@firstoftwo}{%
 \@ifnum{\z@=\pdfoutput}{\@firstoftwo}{\@secondoftwo}%
}{%
 \providecommand\@@startlink[1]{\leavevmode}%
 \providecommand\@@endlink[0]{}%
}{%
 \providecommand\@@startlink[1]{%
  \leavevmode
  \pdfstartlink
   attr{/Border[0 0 1 ]/H/I/C[0 1 1]}%
   user{/Subtype/Link/A<</Type/Action/S/URI/URI(#1)>>}%
  \relax
 }%
 \providecommand\@@endlink[0]{\pdfendlink}%
}%
\providecommand \url  [0]{\begingroup\@sanitize \@url }%
\providecommand \@url [1]{\endgroup\@href {#1}{\urlprefix}}%
\providecommand \urlprefix [0]{URL }%
\providecommand \Eprint[0]{\href }%
\@ifxundefined \urlstyle {%
  \providecommand \doi [1]{doi:\discretionary{}{}{}#1}%
}{%
  \providecommand \doi [0]{doi:\discretionary{}{}{}\begingroup
  \urlstyle{rm}\Url }%
}%
\providecommand \doibase [0]{http://dx.doi.org/}%
\providecommand \Doi[1]{\href{\doibase#1}}%
\providecommand \bibAnnote [3]{%
  \BibitemShut{#1}%
  \begin{quotation}\noindent
    \textsc{Key:}\ #2\\\textsc{Annotation:}\ #3%
  \end{quotation}%
}%
\providecommand \bibAnnoteFile [2]{%
  \IfFileExists{#2}{\bibAnnote {#1} {#2} {\input{#2}}}{}%
}%
\providecommand \typeout [0]{\immediate \write \m@ne }%
\providecommand \selectlanguage [0]{\@gobble}%
\providecommand \bibinfo [0]{\@secondoftwo}%
\providecommand \bibfield [0]{\@secondoftwo}%
\providecommand \translation [1]{[#1]}%
\providecommand \BibitemOpen[0]{}%
\providecommand \bibitemStop [0]{}%
\providecommand \bibitemNoStop [0]{.\EOS\space}%
\providecommand \EOS [0]{\spacefactor3000\relax}%
\providecommand \BibitemShut [1]{\csname bibitem#1\endcsname}%
%</preamble>
\bibitem{YSK+01}%
  \BibitemOpen
  \bibfield{author}{%
  \bibinfo {author} {\bibfnamefont{I.}~\bibnamefont{Yang}}, \bibinfo {author}
  {\bibfnamefont{S.~Y.}\ \bibnamefont{Savrasov}},\ and\ \bibinfo {author}
  {\bibfnamefont{G.}~\bibnamefont{Kotliar}},\ }%
  \bibfield{journal}{%
  \bibinfo {journal} {Phys. Rev. Lett.}\ }%
  \textbf{\bibinfo {volume} {{87}}},\ \bibinfo {pages} {216405} (\bibinfo
  {year} {{2001}})%
  \bibAnnoteFile{NoStop}{YSK+01}%
\bibitem{SM+03}%
  \BibitemOpen
  \bibfield{author}{%
  \bibinfo {author} {\bibfnamefont{A.~B.}\ \bibnamefont{Shick}}\ and\ \bibinfo
  {author} {\bibfnamefont{O.~N.}\ \bibnamefont{Mryasov}},\ }%
  \bibfield{journal}{%
  \bibinfo {journal} {Phys. Rev. B}\ }%
  \textbf{\bibinfo {volume} {67}},\ \bibinfo {pages} {172407} (\bibinfo {year}
  {2003})%
  \bibAnnoteFile{NoStop}{SM+03}%
\bibitem{STK+05}%
  \BibitemOpen
  \bibfield{author}{%
  \bibinfo {author} {\bibfnamefont{S.~Y.}\ \bibnamefont{Savrasov}}, \bibinfo
  {author} {\bibfnamefont{A.}~\bibnamefont{Toropova}}, \bibinfo {author}
  {\bibfnamefont{M.~I.}\ \bibnamefont{Katsnelson}}, \bibinfo {author}
  {\bibfnamefont{A.~I.}\ \bibnamefont{Lichtenstein}}, \bibinfo {author}
  {\bibfnamefont{V.}~\bibnamefont{Antropov}},\ and\ \bibinfo {author}
  {\bibfnamefont{G.}~\bibnamefont{Kotliar}},\ }%
  \bibfield{journal}{%
  \bibinfo {journal} {Z. Kristallogr.}\ }%
  \textbf{\bibinfo {volume} {220}},\ \bibinfo {pages} {473} (\bibinfo {year}
  {2005})%
  \bibAnnoteFile{NoStop}{STK+05}%
\bibitem{SBME10}%
  \BibitemOpen
  \bibfield{author}{%
  \bibinfo {author} {\bibfnamefont{O.}~\bibnamefont{\v{S}ipr}}, \bibinfo
  {author} {\bibfnamefont{S.}~\bibnamefont{Bornemann}}, \bibinfo {author}
  {\bibfnamefont{J.}~\bibnamefont{Min\'ar}},\ and\ \bibinfo {author}
  {\bibfnamefont{H.}~\bibnamefont{Ebert}},\ }%
  \bibfield{journal}{%
  \bibinfo {journal} {Phys. Rev. B}\ }%
  \textbf{\bibinfo {volume} {82}},\ \bibinfo {pages} {174414} (\bibinfo {year}
  {2010})%
  \bibAnnoteFile{NoStop}{SBME10}%
\bibitem{WWF93a}%
  \BibitemOpen
  \bibfield{author}{%
  \bibinfo {author} {\bibfnamefont{D.~S.}\ \bibnamefont{Wang}}, \bibinfo
  {author} {\bibfnamefont{R.}~\bibnamefont{Wu}},\ and\ \bibinfo {author}
  {\bibfnamefont{A.~J.}\ \bibnamefont{Freeman}},\ }%
  \bibfield{journal}{%
  \bibinfo {journal} {Phys. Rev. B}\ }%
  \textbf{\bibinfo {volume} {48}},\ \bibinfo {pages} {15886} (\bibinfo {year}
  {1993})%
  \bibAnnoteFile{NoStop}{WWF93a}%
\bibitem{DKS+94}%
  \BibitemOpen
  \bibfield{author}{%
  \bibinfo {author} {\bibfnamefont{G.~H.~O.}\ \bibnamefont{Daalderop}},
  \bibinfo {author} {\bibfnamefont{P.~J.}\ \bibnamefont{Kelly}},\ and\ \bibinfo
  {author} {\bibfnamefont{M.~F.~H.}\ \bibnamefont{Schuurmans}},\ }%
  \bibfield{journal}{%
  \bibinfo {journal} {Phys. Rev. B}\ }%
  \textbf{\bibinfo {volume} {50}},\ \bibinfo {pages} {9989} (\bibinfo {year}
  {1994})%
  \bibAnnoteFile{NoStop}{DKS+94}%
\bibitem{GC+12}%
  \BibitemOpen
  \bibfield{author}{%
  \bibinfo {author} {\bibfnamefont{F.}~\bibnamefont{Gimbert}}\ and\ \bibinfo
  {author} {\bibfnamefont{L.}~\bibnamefont{Calmels}},\ }%
  \bibfield{journal}{%
  \bibinfo {journal} {Phys. Rev. B}\ }%
  \textbf{\bibinfo {volume} {86}},\ \bibinfo {pages} {184407} (\bibinfo {year}
  {2012})%
  \bibAnnoteFile{NoStop}{GC+12}%
\bibitem{TO+09}%
  \BibitemOpen
  \bibfield{author}{%
  \bibinfo {author} {\bibfnamefont{M.}~\bibnamefont{Tsujikawa}}\ and\ \bibinfo
  {author} {\bibfnamefont{T.}~\bibnamefont{Oda}},\ }%
  \bibfield{journal}{%
  \bibinfo {journal} {Phys. Rev. Lett.}\ }%
  \textbf{\bibinfo {volume} {102}},\ \bibinfo {pages} {247203} (\bibinfo {year}
  {2009})%
  \bibAnnoteFile{NoStop}{TO+09}%
\bibitem{NBL+92}%
  \BibitemOpen
  \bibfield{author}{%
  \bibinfo {author} {\bibfnamefont{L.}~\bibnamefont{Nordstrom}}, \bibinfo
  {author} {\bibfnamefont{M.~S.~S.}\ \bibnamefont{Brooks}},\ and\ \bibinfo
  {author} {\bibfnamefont{B.}~\bibnamefont{Johansson}},\ }%
  \bibfield{journal}{%
  \bibinfo {journal} {J. Phys.: Condens. Matter}\ }%
  \textbf{\bibinfo {volume} {4}},\ \bibinfo {pages} {3261} (\bibinfo {year}
  {1992})%
  \bibAnnoteFile{NoStop}{NBL+92}%
\bibitem{USW96}%
  \BibitemOpen
  \bibfield{author}{%
  \bibinfo {author} {\bibfnamefont{B.}~\bibnamefont{\'Ujfalussy}}, \bibinfo
  {author} {\bibfnamefont{L.}~\bibnamefont{Szunyogh}},\ and\ \bibinfo {author}
  {\bibfnamefont{P.}~\bibnamefont{Weinberger}},\ }%
  \bibfield{journal}{%
  \bibinfo {journal} {Phys. Rev. B}\ }%
  \textbf{\bibinfo {volume} {54}},\ \bibinfo {pages} {9883} (\bibinfo {year}
  {1996})%
  \bibAnnoteFile{NoStop}{USW96}%
\bibitem{HUZ+02}%
  \BibitemOpen
  \bibfield{author}{%
  \bibinfo {author} {\bibfnamefont{R.}~\bibnamefont{Hammerling}}, \bibinfo
  {author} {\bibfnamefont{C.}~\bibnamefont{Uiberacker}}, \bibinfo {author}
  {\bibfnamefont{J.}~\bibnamefont{Zabloudil}}, \bibinfo {author}
  {\bibfnamefont{P.}~\bibnamefont{Weinberger}}, \bibinfo {author}
  {\bibfnamefont{L.}~\bibnamefont{Szunyogh}},\ and\ \bibinfo {author}
  {\bibfnamefont{J.}~\bibnamefont{Kirschner}},\ }%
  \bibfield{journal}{%
  \bibinfo {journal} {Phys. Rev. B}\ }%
  \textbf{\bibinfo {volume} {66}},\ \bibinfo {pages} {052402} (\bibinfo {year}
  {2002})%
  \bibAnnoteFile{NoStop}{HUZ+02}%
\bibitem{USBW96}%
  \BibitemOpen
  \bibfield{author}{%
  \bibinfo {author} {\bibfnamefont{B.}~\bibnamefont{\'Ujfalussy}}, \bibinfo
  {author} {\bibfnamefont{L.}~\bibnamefont{Szunyogh}}, \bibinfo {author}
  {\bibfnamefont{P.}~\bibnamefont{Bruno}},\ and\ \bibinfo {author}
  {\bibfnamefont{P.}~\bibnamefont{Weinberger}},\ }%
  \bibfield{journal}{%
  \bibinfo {journal} {Phys. Rev. Lett.}\ }%
  \textbf{\bibinfo {volume} {77}},\ \bibinfo {pages} {1805} (\bibinfo {year}
  {1996})%
  \bibAnnoteFile{NoStop}{USBW96}%
\bibitem{CE+97}%
  \BibitemOpen
  \bibfield{author}{%
  \bibinfo {author} {\bibfnamefont{M.}~\bibnamefont{Cinal}}\ and\ \bibinfo
  {author} {\bibfnamefont{D.~M.}\ \bibnamefont{Edwards}},\ }%
  \bibfield{journal}{%
  \bibinfo {journal} {Phys. Rev. B}\ }%
  \textbf{\bibinfo {volume} {55}},\ \bibinfo {pages} {3636} (\bibinfo {year}
  {1997})%
  \bibAnnoteFile{NoStop}{CE+97}%
\bibitem{DDP03}%
  \BibitemOpen
  \bibfield{author}{%
  \bibinfo {author} {\bibfnamefont{J.}~\bibnamefont{Dorantes-D\'avila}},
  \bibinfo {author} {\bibfnamefont{H.}~\bibnamefont{Dreyss\'e}},\ and\ \bibinfo
  {author} {\bibfnamefont{G.~M.}\ \bibnamefont{Pastor}},\ }%
  \bibfield{journal}{%
  \bibinfo {journal} {Phys. Rev. Lett.}\ }%
  \textbf{\bibinfo {volume} {91}},\ \bibinfo {pages} {197206} (\bibinfo {year}
  {2003})%
  \bibAnnoteFile{NoStop}{DDP03}%
\bibitem{BES+05}%
  \BibitemOpen
  \bibfield{author}{%
  \bibinfo {author} {\bibfnamefont{T.}~\bibnamefont{Burkert}}, \bibinfo
  {author} {\bibfnamefont{O.}~\bibnamefont{Eriksson}}, \bibinfo {author}
  {\bibfnamefont{S.~I.}\ \bibnamefont{Simak}}, \bibinfo {author}
  {\bibfnamefont{A.~V.}\ \bibnamefont{Ruban}}, \bibinfo {author}
  {\bibfnamefont{B.}~\bibnamefont{Sanyal}}, \bibinfo {author}
  {\bibfnamefont{L.}~\bibnamefont{Nordstr\"om}},\ and\ \bibinfo {author}
  {\bibfnamefont{J.~M.}\ \bibnamefont{Wills}},\ }%
  \bibfield{journal}{%
  \bibinfo {journal} {Phys. Rev. B}\ }%
  \textbf{\bibinfo {volume} {71}},\ \bibinfo {pages} {134411} (\bibinfo {year}
  {2005})%
  \bibAnnoteFile{NoStop}{BES+05}%
\bibitem{KSF+06}%
  \BibitemOpen
  \bibfield{author}{%
  \bibinfo {author} {\bibfnamefont{M.}~\bibnamefont{Komelj}}, \bibinfo {author}
  {\bibfnamefont{D.}~\bibnamefont{Steiauf}},\ and\ \bibinfo {author}
  {\bibfnamefont{M.}~\bibnamefont{F\"{a}hnle}},\ }%
  \bibfield{journal}{%
  \bibinfo {journal} {Phys. Rev. B}\ }%
  \textbf{\bibinfo {volume} {73}},\ \bibinfo {pages} {134428} (\bibinfo {year}
  {2006})%
  \bibAnnoteFile{NoStop}{KSF+06}%
\bibitem{BLD+10}%
  \BibitemOpen
  \bibfield{author}{%
  \bibinfo {author} {\bibfnamefont{P.}~\bibnamefont{B{\l}o\'{n}ski}}, \bibinfo
  {author} {\bibfnamefont{A.}~\bibnamefont{Lehnert}}, \bibinfo {author}
  {\bibfnamefont{S.}~\bibnamefont{Dennler}}, \bibinfo {author}
  {\bibfnamefont{S.}~\bibnamefont{Rusponi}}, \bibinfo {author}
  {\bibfnamefont{M.}~\bibnamefont{Etzkorn}}, \bibinfo {author}
  {\bibfnamefont{G.}~\bibnamefont{Moulas}}, \bibinfo {author}
  {\bibfnamefont{P.}~\bibnamefont{Bencok}}, \bibinfo {author}
  {\bibfnamefont{P.}~\bibnamefont{Gambardella}}, \bibinfo {author}
  {\bibfnamefont{H.}~\bibnamefont{Brune}},\ and\ \bibinfo {author}
  {\bibfnamefont{J.}~\bibnamefont{Hafner}},\ }%
  \bibfield{journal}{%
  \bibinfo {journal} {Phys. Rev. B}\ }%
  \textbf{\bibinfo {volume} {81}},\ \bibinfo {pages} {104426} (\bibinfo {year}
  {2010})%
  \bibAnnoteFile{NoStop}{BLD+10}%
\bibitem{BSM+12}%
  \BibitemOpen
  \bibfield{author}{%
  \bibinfo {author} {\bibfnamefont{S.}~\bibnamefont{Bornemann}}, \bibinfo
  {author} {\bibfnamefont{O.}~\bibnamefont{\v{S}ipr}}, \bibinfo {author}
  {\bibfnamefont{S.}~\bibnamefont{Mankovsky}}, \bibinfo {author}
  {\bibfnamefont{S.}~\bibnamefont{Polesya}}, \bibinfo {author}
  {\bibfnamefont{J.~B.}\ \bibnamefont{Staunton}}, \bibinfo {author}
  {\bibfnamefont{W.}~\bibnamefont{Wurth}}, \bibinfo {author}
  {\bibfnamefont{H.}~\bibnamefont{Ebert}},\ and\ \bibinfo {author}
  {\bibfnamefont{J.}~\bibnamefont{Min\'{a}r}},\ }%
  \bibfield{journal}{%
  \bibinfo {journal} {Phys. Rev. B}\ }%
  \textbf{\bibinfo {volume} {86}},\ \bibinfo {pages} {104436} (\bibinfo {year}
  {2012})%
  \bibAnnoteFile{NoStop}{BSM+12}%
\bibitem{APS+12}%
  \BibitemOpen
  \bibfield{author}{%
  \bibinfo {author} {\bibfnamefont{C.~J.}\ \bibnamefont{Aas}}, \bibinfo
  {author} {\bibfnamefont{K.}~\bibnamefont{Palot\'{a}s}}, \bibinfo {author}
  {\bibfnamefont{L.}~\bibnamefont{Szunyogh}},\ and\ \bibinfo {author}
  {\bibfnamefont{R.~W.}\ \bibnamefont{Chantrell}},\ }%
  \bibfield{journal}{%
  \bibinfo {journal} {J. Phys.: Condens. Matter}\ }%
  \textbf{\bibinfo {volume} {24}},\ \bibinfo {pages} {406001} (\bibinfo {year}
  {2012})%
  \bibAnnoteFile{NoStop}{APS+12}%
\bibitem{DVS+08}%
  \BibitemOpen
  \bibfield{author}{%
  \bibinfo {author} {\bibfnamefont{C.-G.}\ \bibnamefont{Duan}}, \bibinfo
  {author} {\bibfnamefont{J.~P.}\ \bibnamefont{Velev}}, \bibinfo {author}
  {\bibfnamefont{R.~F.}\ \bibnamefont{Sabirianov}}, \bibinfo {author}
  {\bibfnamefont{Z.}~\bibnamefont{Zhu}}, \bibinfo {author}
  {\bibfnamefont{J.}~\bibnamefont{Chu}}, \bibinfo {author}
  {\bibfnamefont{S.~S.}\ \bibnamefont{Jaswal}},\ and\ \bibinfo {author}
  {\bibfnamefont{E.~Y.}\ \bibnamefont{Tsymbal}},\ }%
  \bibfield{journal}{%
  \bibinfo {journal} {Phys. Rev. Lett.}\ }%
  \textbf{\bibinfo {volume} {101}},\ \bibinfo {pages} {137201} (\bibinfo {year}
  {2008})%
  \bibAnnoteFile{NoStop}{DVS+08}%
\bibitem{USW97}%
  \BibitemOpen
  \bibfield{author}{%
  \bibinfo {author} {\bibfnamefont{B.}~\bibnamefont{\'Ujfalussy}}, \bibinfo
  {author} {\bibfnamefont{L.}~\bibnamefont{Szunyogh}},\ and\ \bibinfo {author}
  {\bibfnamefont{P.}~\bibnamefont{Weinberger}},\ }%
  in\ \emph{\bibinfo {booktitle} {Properties of Complex Inorganic Solids}},\
  \bibinfo {editor} {edited by\ \bibinfo {editor}
  {\bibfnamefont{A.}~\bibnamefont{Gonis}}, \bibinfo {editor}
  {\bibfnamefont{A.}~\bibnamefont{Meike}},\ and\ \bibinfo {editor}
  {\bibfnamefont{P.~E.~A.}\ \bibnamefont{Turchi}}}\ (\bibinfo {publisher}
  {Plenum Press},\ \bibinfo {address} {New York},\ \bibinfo {year} {1997})\ p.\
  \bibinfo {pages} {181}%
  \bibAnnoteFile{NoStop}{USW97}%
\bibitem{BRB+06}%
  \BibitemOpen
  \bibfield{author}{%
  \bibinfo {author} {\bibfnamefont{S.}~\bibnamefont{Baud}}, \bibinfo {author}
  {\bibfnamefont{C.}~\bibnamefont{Ramseyer}}, \bibinfo {author}
  {\bibfnamefont{G.}~\bibnamefont{Bihlmayer}},\ and\ \bibinfo {author}
  {\bibfnamefont{S.}~\bibnamefont{Bl\"ugel}},\ }%
  \bibfield{journal}{%
  \bibinfo {journal} {Phys. Rev. B}\ }%
  \textbf{\bibinfo {volume} {73}},\ \bibinfo {pages} {104427} (\bibinfo {year}
  {2006})%
  \bibAnnoteFile{NoStop}{BRB+06}%
\bibitem{SF+11}%
  \BibitemOpen
  \bibfield{author}{%
  \bibinfo {author} {\bibfnamefont{S.}~\bibnamefont{Subkow}}\ and\ \bibinfo
  {author} {\bibfnamefont{M.}~\bibnamefont{F\"ahnle}},\ }%
  \bibfield{journal}{%
  \bibinfo {journal} {Phys. Rev. B}\ }%
  \textbf{\bibinfo {volume} {84}},\ \bibinfo {pages} {054443} (\bibinfo {year}
  {2011})%
  \bibAnnoteFile{NoStop}{SF+11}%
\bibitem{KSM+11}%
  \BibitemOpen
  \bibfield{author}{%
  \bibinfo {author} {\bibfnamefont{S.}~\bibnamefont{Khmelevskyi}}, \bibinfo
  {author} {\bibfnamefont{A.~B.}\ \bibnamefont{Shick}},\ and\ \bibinfo {author}
  {\bibfnamefont{P.}~\bibnamefont{Mohn}},\ }%
  \bibfield{journal}{%
  \bibinfo {journal} {Phys. Rev. B}\ }%
  \textbf{\bibinfo {volume} {83}},\ \bibinfo {pages} {224419} (\bibinfo {year}
  {2011})%
  \bibAnnoteFile{NoStop}{KSM+11}%
\bibitem{SMO+08}%
  \BibitemOpen
  \bibfield{author}{%
  \bibinfo {author} {\bibfnamefont{A.~B.}\ \bibnamefont{Shick}}, \bibinfo
  {author} {\bibfnamefont{F.}~\bibnamefont{M\'{a}ca}}, \bibinfo {author}
  {\bibfnamefont{M.}~\bibnamefont{Ondr\'{a}\v{c}ek}}, \bibinfo {author}
  {\bibfnamefont{O.~N.}\ \bibnamefont{Mryasov}},\ and\ \bibinfo {author}
  {\bibfnamefont{T.}~\bibnamefont{Jungwirth}},\ }%
  \bibfield{journal}{%
  \bibinfo {journal} {Phys. Rev. B}\ }%
  \textbf{\bibinfo {volume} {78}},\ \bibinfo {pages} {054413} (\bibinfo {year}
  {2008})%
  \bibAnnoteFile{NoStop}{SMO+08}%
\bibitem{SDM95}%
  \BibitemOpen
  \bibfield{author}{%
  \bibinfo {author} {\bibfnamefont{I.~V.}\ \bibnamefont{Solovyev}}, \bibinfo
  {author} {\bibfnamefont{P.~H.}\ \bibnamefont{Dederichs}},\ and\ \bibinfo
  {author} {\bibfnamefont{I.}~\bibnamefont{Mertig}},\ }%
  \bibfield{journal}{%
  \bibinfo {journal} {Phys. Rev. B}\ }%
  \textbf{\bibinfo {volume} {52}},\ \bibinfo {pages} {13419} (\bibinfo {year}
  {1995})%
  \bibAnnoteFile{NoStop}{SDM95}%
\bibitem{RKF+01}%
  \BibitemOpen
  \bibfield{author}{%
  \bibinfo {author} {\bibfnamefont{P.}~\bibnamefont{Ravindran}}, \bibinfo
  {author} {\bibfnamefont{A.}~\bibnamefont{Kjekshus}}, \bibinfo {author}
  {\bibfnamefont{H.}~\bibnamefont{Fjellv{\aa}g}}, \bibinfo {author}
  {\bibfnamefont{P.}~\bibnamefont{James}}, \bibinfo {author}
  {\bibfnamefont{L.}~\bibnamefont{Nordstr\"om}}, \bibinfo {author}
  {\bibfnamefont{B.}~\bibnamefont{Johansson}},\ and\ \bibinfo {author}
  {\bibfnamefont{O.}~\bibnamefont{Eriksson}},\ }%
  \bibfield{journal}{%
  \bibinfo {journal} {Phys. Rev. B}\ }%
  \textbf{\bibinfo {volume} {63}},\ \bibinfo {pages} {144409} (\bibinfo {year}
  {2001})%
  \bibAnnoteFile{NoStop}{RKF+01}%
\bibitem{Cin+01}%
  \BibitemOpen
  \bibfield{author}{%
  \bibinfo {author} {\bibfnamefont{M.}~\bibnamefont{Cinal}},\ }%
  \bibfield{journal}{%
  \bibinfo {journal} {J. Phys.: Condens. Matter}\ }%
  \textbf{\bibinfo {volume} {13}},\ \bibinfo {pages} {901} (\bibinfo {year}
  {2001})%
  \bibAnnoteFile{NoStop}{Cin+01}%
\bibitem{SML+09}%
  \BibitemOpen
  \bibfield{author}{%
  \bibinfo {author} {\bibfnamefont{A.~B.}\ \bibnamefont{Shick}}, \bibinfo
  {author} {\bibfnamefont{F.}~\bibnamefont{M\'{a}ca}},\ and\ \bibinfo {author}
  {\bibfnamefont{A.~I.}\ \bibnamefont{Lichtenstein}},\ }%
  \bibfield{journal}{%
  \bibinfo {journal} {J. Appl. Phys.}\ }%
  \textbf{\bibinfo {volume} {105}},\ \bibinfo {pages} {07C309} (\bibinfo {year}
  {2009})%
  \bibAnnoteFile{NoStop}{SML+09}%
\bibitem{SF+09}%
  \BibitemOpen
  \bibfield{author}{%
  \bibinfo {author} {\bibfnamefont{S.}~\bibnamefont{Subkow}}\ and\ \bibinfo
  {author} {\bibfnamefont{M.}~\bibnamefont{F\"{a}hnle}},\ }%
  \bibfield{journal}{%
  \bibinfo {journal} {Phys. Rev. B}\ }%
  \textbf{\bibinfo {volume} {80}},\ \bibinfo {pages} {212404} (\bibinfo {year}
  {2009})%
  \bibAnnoteFile{NoStop}{SF+09}%
\bibitem{MLR+08}%
  \BibitemOpen
  \bibfield{author}{%
  \bibinfo {author} {\bibfnamefont{G.}~\bibnamefont{Moulas}}, \bibinfo {author}
  {\bibfnamefont{A.}~\bibnamefont{Lehnert}}, \bibinfo {author}
  {\bibfnamefont{S.}~\bibnamefont{Rusponi}}, \bibinfo {author}
  {\bibfnamefont{J.}~\bibnamefont{Zabloudil}}, \bibinfo {author}
  {\bibfnamefont{C.}~\bibnamefont{Etz}}, \bibinfo {author}
  {\bibfnamefont{S.}~\bibnamefont{Ouazi}}, \bibinfo {author}
  {\bibfnamefont{M.}~\bibnamefont{Etzkorn}}, \bibinfo {author}
  {\bibfnamefont{P.}~\bibnamefont{Bencok}}, \bibinfo {author}
  {\bibfnamefont{P.}~\bibnamefont{Gambardella}}, \bibinfo {author}
  {\bibfnamefont{P.}~\bibnamefont{Weinberger}},\ and\ \bibinfo {author}
  {\bibfnamefont{H.}~\bibnamefont{Brune}},\ }%
  \bibfield{journal}{%
  \bibinfo {journal} {Phys. Rev. B}\ }%
  \textbf{\bibinfo {volume} {78}},\ \bibinfo {pages} {214424} (\bibinfo {year}
  {2008})%
  \bibAnnoteFile{NoStop}{MLR+08}%
\bibitem{VWN80}%
  \BibitemOpen
  \bibfield{author}{%
  \bibinfo {author} {\bibfnamefont{S.~H.}\ \bibnamefont{Vosko}}, \bibinfo
  {author} {\bibfnamefont{L.}~\bibnamefont{Wilk}},\ and\ \bibinfo {author}
  {\bibfnamefont{M.}~\bibnamefont{Nusair}},\ }%
  \bibfield{journal}{%
  \bibinfo {journal} {Can. J. Phys.}\ }%
  \textbf{\bibinfo {volume} {58}},\ \bibinfo {pages} {1200} (\bibinfo {year}
  {1980})%
  \bibAnnoteFile{NoStop}{VWN80}%
\bibitem{EKM11}%
  \BibitemOpen
  \bibfield{author}{%
  \bibinfo {author} {\bibfnamefont{H.}~\bibnamefont{Ebert}}, \bibinfo {author}
  {\bibfnamefont{D.}~\bibnamefont{K\"odderitzsch}},\ and\ \bibinfo {author}
  {\bibnamefont{Min\'{a}r}},\ }%
  \bibfield{journal}{%
  \bibinfo {journal} {Rep. Prog. Phys.}\ }%
  \textbf{\bibinfo {volume} {74}},\ \bibinfo {pages} {096501} (\bibinfo {year}
  {2011})%
  \bibAnnoteFile{NoStop}{EKM11}%
\bibitem{tbkkr-code}%
  \BibitemOpen
  \bibfield{author}{%
  \bibinfo {author} {\bibfnamefont{H.}~\bibnamefont{Ebert}}\ and\ \bibinfo
  {author} {\bibfnamefont{R.}~\bibnamefont{Zeller}},\ }%
  \emph{\bibinfo {title} {The {\sc spr-tb-kkr} package}},\ \bibinfo {address}
  {\url{http://olymp.cup.uni-muenchen.de}} (\bibinfo {year} {2006})%
  \bibAnnoteFile{NoStop}{tbkkr-code}%
\bibitem{DWW+88}%
  \BibitemOpen
  \bibfield{author}{%
  \bibinfo {author} {\bibfnamefont{J.~W.}\ \bibnamefont{Davenport}}, \bibinfo
  {author} {\bibfnamefont{R.~E.}\ \bibnamefont{Watson}},\ and\ \bibinfo
  {author} {\bibfnamefont{M.}~\bibnamefont{Weinert}},\ }%
  \bibfield{journal}{%
  \bibinfo {journal} {Phys. Rev. B}\ }%
  \textbf{\bibinfo {volume} {37}},\ \bibinfo {pages} {9985} (\bibinfo {year}
  {1988})%
  \bibAnnoteFile{NoStop}{DWW+88}%
\bibitem{CRC+07}%
  \BibitemOpen
  \bibfield{author}{%
  \bibinfo {author} {\bibfnamefont{D.~R.}\ \bibnamefont{Lide}},\ }%
  \emph{\bibinfo {title} {CRC Handbook of Chemistry and Physics, 88th
  Edition}}\ (\bibinfo {publisher} {Taylor and Francis},\ \bibinfo {address}
  {Boca Raton},\ \bibinfo {year} {2007})%
  \bibAnnoteFile{NoStop}{CRC+07}%
\bibitem{MDV+82}%
  \BibitemOpen
  \bibfield{author}{%
  \bibinfo {author} {\bibfnamefont{A.~H.}\ \bibnamefont{MacDonald}}, \bibinfo
  {author} {\bibfnamefont{J.~M.}\ \bibnamefont{Daams}}, \bibinfo {author}
  {\bibfnamefont{S.~H.}\ \bibnamefont{Vosko}},\ and\ \bibinfo {author}
  {\bibfnamefont{D.~D.}\ \bibnamefont{Koelling}},\ }%
  \bibfield{journal}{%
  \bibinfo {journal} {Phys. Rev. B}\ }%
  \textbf{\bibinfo {volume} {25}},\ \bibinfo {pages} {713} (\bibinfo {year}
  {1982})%
  \bibAnnoteFile{NoStop}{MDV+82}%
\bibitem{SLP+93}%
  \BibitemOpen
  \bibfield{author}{%
  \bibinfo {author} {\bibfnamefont{V.~I.}\ \bibnamefont{Smelyansky}}, \bibinfo
  {author} {\bibfnamefont{M.~J.~G.}\ \bibnamefont{Lee}},\ and\ \bibinfo
  {author} {\bibfnamefont{J.~M.}\ \bibnamefont{Perz}},\ }%
  \bibfield{journal}{%
  \bibinfo {journal} {J. Phys.: Condens. Matter}\ }%
  \textbf{\bibinfo {volume} {5}},\ \bibinfo {pages} {6061} (\bibinfo {year}
  {1993})%
  \bibAnnoteFile{NoStop}{SLP+93}%
\bibitem{SV+80}%
  \BibitemOpen
  \bibfield{author}{%
  \bibinfo {author} {\bibfnamefont{W.}~\bibnamefont{S\"{a}nger}}\ and\ \bibinfo
  {author} {\bibfnamefont{J.}~\bibnamefont{Voitl\"{a}nder}},\ }%
  \bibfield{journal}{%
  \bibinfo {journal} {Z. Physik B}\ }%
  \textbf{\bibinfo {volume} {38}},\ \bibinfo {pages} {133} (\bibinfo {year}
  {1980})%
  \bibAnnoteFile{NoStop}{SV+80}%
\bibitem{PVF+10}%
  \BibitemOpen
  \bibfield{author}{%
  \bibinfo {author} {\bibfnamefont{A.}~\bibnamefont{Povzner}}, \bibinfo
  {author} {\bibfnamefont{A.}~\bibnamefont{Volkov}},\ and\ \bibinfo {author}
  {\bibfnamefont{A.}~\bibnamefont{Filanovich}},\ }%
  \bibfield{journal}{%
  \bibinfo {journal} {Physics of the Solid State}\ }%
  \textbf{\bibinfo {volume} {52}},\ \bibinfo {pages} {2012} (\bibinfo {year}
  {2010})%
  \bibAnnoteFile{NoStop}{PVF+10}%
\bibitem{ZDU+95}%
  \BibitemOpen
  \bibfield{author}{%
  \bibinfo {author} {\bibfnamefont{R.}~\bibnamefont{Zeller}}, \bibinfo {author}
  {\bibfnamefont{P.~H.}\ \bibnamefont{Dederichs}}, \bibinfo {author}
  {\bibfnamefont{B.}~\bibnamefont{\'Ujfalussy}}, \bibinfo {author}
  {\bibfnamefont{L.}~\bibnamefont{Szunyogh}},\ and\ \bibinfo {author}
  {\bibfnamefont{P.}~\bibnamefont{Weinberger}},\ }%
  \bibfield{journal}{%
  \bibinfo {journal} {Phys. Rev. B}\ }%
  \textbf{\bibinfo {volume} {52}},\ \bibinfo {pages} {8807} (\bibinfo {year}
  {1995})%
  \bibAnnoteFile{NoStop}{ZDU+95}%
\bibitem{BMP+05}%
  \BibitemOpen
  \bibfield{author}{%
  \bibinfo {author} {\bibfnamefont{S.}~\bibnamefont{Bornemann}}, \bibinfo
  {author} {\bibfnamefont{J.}~\bibnamefont{Min\'{a}r}}, \bibinfo {author}
  {\bibfnamefont{S.}~\bibnamefont{Polesya}}, \bibinfo {author}
  {\bibfnamefont{S.}~\bibnamefont{Mankovsky}}, \bibinfo {author}
  {\bibfnamefont{H.}~\bibnamefont{Ebert}},\ and\ \bibinfo {author}
  {\bibfnamefont{O.}~\bibnamefont{\v{S}ipr}},\ }%
  \bibfield{journal}{%
  \bibinfo {journal} {Phase Transitions}\ }%
  \textbf{\bibinfo {volume} {78}},\ \bibinfo {pages} {701} (\bibinfo {year}
  {2005})%
  \bibAnnoteFile{NoStop}{BMP+05}%
\bibitem{Bir66}%
  \BibitemOpen
  \bibfield{author}{%
  \bibinfo {author} {\bibfnamefont{R.~R.}\ \bibnamefont{Birss}},\ }%
  \emph{\bibinfo {title} {Symmetry and Magnetism}},\ \bibinfo {series}
  {Selected Topics in Solid State Physics}, Vol.~\bibinfo {volume} {3}\
  (\bibinfo {publisher} {North-Holland},\ \bibinfo {address} {Amsterdam},\
  \bibinfo {year} {1966})%
  \bibAnnoteFile{NoStop}{Bir66}%
\bibitem{WWW+96}%
  \BibitemOpen
  \bibfield{author}{%
  \bibinfo {author} {\bibfnamefont{X.~D.}\ \bibnamefont{Wang}}, \bibinfo
  {author} {\bibfnamefont{R.}~\bibnamefont{Wu}}, \bibinfo {author}
  {\bibfnamefont{D.~S.}\ \bibnamefont{Wang}},\ and\ \bibinfo {author}
  {\bibfnamefont{A.~J.}\ \bibnamefont{Freeman}},\ }%
  \bibfield{journal}{%
  \bibinfo {journal} {Phys. Rev. B}\ }%
  \textbf{\bibinfo {volume} {54}},\ \bibinfo {pages} {61} (\bibinfo {year}
  {1996})%
  \bibAnnoteFile{NoStop}{WWW+96}%
\bibitem{SSB+06}%
  \BibitemOpen
  \bibfield{author}{%
  \bibinfo {author} {\bibfnamefont{J.~B.}\ \bibnamefont{Staunton}}, \bibinfo
  {author} {\bibfnamefont{L.}~\bibnamefont{Szunyogh}}, \bibinfo {author}
  {\bibfnamefont{A.}~\bibnamefont{Buruzs}}, \bibinfo {author}
  {\bibfnamefont{B.~L.}\ \bibnamefont{Gyorffy}}, \bibinfo {author}
  {\bibfnamefont{S.}~\bibnamefont{Ostanin}},\ and\ \bibinfo {author}
  {\bibfnamefont{L.}~\bibnamefont{Udvardi}},\ }%
  \bibfield{journal}{%
  \bibinfo {journal} {Phys. Rev. B}\ }%
  \textbf{\bibinfo {volume} {74}},\ \bibinfo {pages} {144411} (\bibinfo {year}
  {2006})%
  \bibAnnoteFile{NoStop}{SSB+06}%
\bibitem{BMB+12}%
  \BibitemOpen
  \bibfield{author}{%
  \bibinfo {author} {\bibfnamefont{S.}~\bibnamefont{Bornemann}}, \bibinfo
  {author} {\bibfnamefont{J.}~\bibnamefont{Min\'{a}r}}, \bibinfo {author}
  {\bibfnamefont{J.}~\bibnamefont{Braun}}, \bibinfo {author}
  {\bibfnamefont{D.}~\bibnamefont{Koedderitzsch}},\ and\ \bibinfo {author}
  {\bibfnamefont{H.}~\bibnamefont{Ebert}},\ }%
  \bibfield{journal}{%
  \bibinfo {journal} {Solid State Commun.}\ }%
  \textbf{\bibinfo {volume} {152}},\ \bibinfo {pages} {85} (\bibinfo {year}
  {2012})%
  \bibAnnoteFile{NoStop}{BMB+12}%
\bibitem{EFVG96}%
  \BibitemOpen
  \bibfield{author}{%
  \bibinfo {author} {\bibfnamefont{H.}~\bibnamefont{Ebert}}, \bibinfo {author}
  {\bibfnamefont{H.}~\bibnamefont{Freyer}}, \bibinfo {author}
  {\bibfnamefont{A.}~\bibnamefont{Vernes}},\ and\ \bibinfo {author}
  {\bibfnamefont{G.-Y.}\ \bibnamefont{Guo}},\ }%
  \bibfield{journal}{%
  \bibinfo {journal} {Phys. Rev. B}\ }%
  \textbf{\bibinfo {volume} {53}},\ \bibinfo {pages} {7721} (\bibinfo {year}
  {1996})%
  \bibAnnoteFile{NoStop}{EFVG96}%
\bibitem{KH77}%
  \BibitemOpen
  \bibfield{author}{%
  \bibinfo {author} {\bibfnamefont{D.~D.}\ \bibnamefont{Koelling}}\ and\
  \bibinfo {author} {\bibfnamefont{B.~N.}\ \bibnamefont{Harmon}},\ }%
  \bibfield{journal}{%
  \bibinfo {journal} {J. Phys. C: Solid State Phys.}\ }%
  \textbf{\bibinfo {volume} {10}},\ \bibinfo {pages} {3107} (\bibinfo {year}
  {1977})%
  \bibAnnoteFile{NoStop}{KH77}%
\bibitem{MV94}%
  \BibitemOpen
  \bibfield{author}{%
  \bibinfo {author} {\bibfnamefont{J.~M.}\ \bibnamefont{MacLaren}}\ and\
  \bibinfo {author} {\bibfnamefont{R.~H.}\ \bibnamefont{Victora}},\ }%
  \bibfield{journal}{%
  \bibinfo {journal} {J. Appl. Phys.}\ }%
  \textbf{\bibinfo {volume} {76}},\ \bibinfo {pages} {6069} (\bibinfo {year}
  {1994})%
  \bibAnnoteFile{NoStop}{MV94}%
\bibitem{EVB99}%
  \BibitemOpen
  \bibfield{author}{%
  \bibinfo {author} {\bibfnamefont{H.}~\bibnamefont{Ebert}}, \bibinfo {author}
  {\bibfnamefont{A.}~\bibnamefont{Vernes}},\ and\ \bibinfo {author}
  {\bibfnamefont{J.}~\bibnamefont{Banhart}},\ }%
  \bibfield{journal}{%
  \bibinfo {journal} {Solid State Commun.}\ }%
  \textbf{\bibinfo {volume} {113}},\ \bibinfo {pages} {103} (\bibinfo {year}
  {2000})%
  \bibAnnoteFile{NoStop}{EVB99}%
\bibitem{MEG+01}%
  \BibitemOpen
  \bibfield{author}{%
  \bibinfo {author} {\bibfnamefont{J.}~\bibnamefont{Min\'ar}}, \bibinfo
  {author} {\bibfnamefont{H.}~\bibnamefont{Ebert}}, \bibinfo {author}
  {\bibfnamefont{G.}~\bibnamefont{Ghiringhelli}}, \bibinfo {author}
  {\bibfnamefont{O.}~\bibnamefont{Tjernberg}}, \bibinfo {author}
  {\bibfnamefont{N.~B.}\ \bibnamefont{Brookes}},\ and\ \bibinfo {author}
  {\bibfnamefont{L.~H.}\ \bibnamefont{Tjeng}},\ }%
  \bibfield{journal}{%
  \bibinfo {journal} {Phys. Rev. B}\ }%
  \textbf{\bibinfo {volume} {63}},\ \bibinfo {pages} {144421} (\bibinfo {year}
  {2001})%
  \bibAnnoteFile{NoStop}{MEG+01}%
\bibitem{PEP+05}%
  \BibitemOpen
  \bibfield{author}{%
  \bibinfo {author} {\bibfnamefont{V.}~\bibnamefont{Popescu}}, \bibinfo
  {author} {\bibfnamefont{H.}~\bibnamefont{Ebert}}, \bibinfo {author}
  {\bibfnamefont{N.}~\bibnamefont{Papanikolaou}}, \bibinfo {author}
  {\bibfnamefont{R.}~\bibnamefont{Zeller}},\ and\ \bibinfo {author}
  {\bibfnamefont{P.~H.}\ \bibnamefont{Dederichs}},\ }%
  \bibfield{journal}{%
  \bibinfo {journal} {Phys. Rev. B}\ }%
  \textbf{\bibinfo {volume} {72}},\ \bibinfo {pages} {184427} (\bibinfo {year}
  {2005})%
  \bibAnnoteFile{NoStop}{PEP+05}%
\bibitem{MKWE13}%
  \BibitemOpen
  \bibfield{author}{%
  \bibinfo {author} {\bibfnamefont{S.}~\bibnamefont{Mankovsky}}, \bibinfo
  {author} {\bibfnamefont{D.}~\bibnamefont{K\"odderitzsch}}, \bibinfo {author}
  {\bibfnamefont{G.}~\bibnamefont{Woltersdorf}},\ and\ \bibinfo {author}
  {\bibfnamefont{H.}~\bibnamefont{Ebert}},\ }%
  \bibfield{journal}{%
  \bibinfo {journal} {Phys. Rev. B}\ }%
  \textbf{\bibinfo {volume} {87}},\ \bibinfo {pages} {014430} (\bibinfo {year}
  {2013})%
  \bibAnnoteFile{NoStop}{MKWE13}%
\bibitem{PMS+10}%
  \BibitemOpen
  \bibfield{author}{%
  \bibinfo {author} {\bibfnamefont{S.}~\bibnamefont{Polesya}}, \bibinfo
  {author} {\bibfnamefont{S.}~\bibnamefont{Mankovsky}}, \bibinfo {author}
  {\bibfnamefont{O.}~\bibnamefont{\v{S}ipr}}, \bibinfo {author}
  {\bibfnamefont{W.}~\bibnamefont{Meindl}}, \bibinfo {author}
  {\bibfnamefont{C.}~\bibnamefont{Strunk}},\ and\ \bibinfo {author}
  {\bibfnamefont{H.}~\bibnamefont{Ebert}},\ }%
  \bibfield{journal}{%
  \bibinfo {journal} {Phys. Rev. B}\ }%
  \textbf{\bibinfo {volume} {82}},\ \bibinfo {pages} {214409} (\bibinfo {year}
  {2010})%
  \bibAnnoteFile{NoStop}{PMS+10}%
\bibitem{SBM+07}%
  \BibitemOpen
  \bibfield{author}{%
  \bibinfo {author} {\bibfnamefont{O.}~\bibnamefont{\v{S}ipr}}, \bibinfo
  {author} {\bibfnamefont{S.}~\bibnamefont{Bornemann}}, \bibinfo {author}
  {\bibfnamefont{J.}~\bibnamefont{Min\'ar}}, \bibinfo {author}
  {\bibfnamefont{S.}~\bibnamefont{Polesya}}, \bibinfo {author}
  {\bibfnamefont{V.}~\bibnamefont{Popescu}}, \bibinfo {author}
  {\bibfnamefont{A.}~\bibnamefont{\v{S}im{\accent23 u}nek}},\ and\ \bibinfo
  {author} {\bibfnamefont{H.}~\bibnamefont{Ebert}},\ }%
  \bibfield{journal}{%
  \bibinfo {journal} {J. Phys.: Condens. Matter}\ }%
  \textbf{\bibinfo {volume} {19}},\ \bibinfo {pages} {096203} (\bibinfo {year}
  {2007})%
  \bibAnnoteFile{NoStop}{SBM+07}%
\bibitem{LDB+10}%
  \BibitemOpen
  \bibfield{author}{%
  \bibinfo {author} {\bibfnamefont{A.}~\bibnamefont{Lehnert}}, \bibinfo
  {author} {\bibfnamefont{S.}~\bibnamefont{Dennler}}, \bibinfo {author}
  {\bibfnamefont{P.}~\bibnamefont{B{\l}o\'{n}ski}}, \bibinfo {author}
  {\bibfnamefont{S.}~\bibnamefont{Rusponi}}, \bibinfo {author}
  {\bibfnamefont{M.}~\bibnamefont{Etzkorn}}, \bibinfo {author}
  {\bibfnamefont{G.}~\bibnamefont{Moulas}}, \bibinfo {author}
  {\bibfnamefont{P.}~\bibnamefont{Bencok}}, \bibinfo {author}
  {\bibfnamefont{P.}~\bibnamefont{Gambardella}}, \bibinfo {author}
  {\bibfnamefont{H.}~\bibnamefont{Brune}},\ and\ \bibinfo {author}
  {\bibfnamefont{J.}~\bibnamefont{Hafner}},\ }%
  \bibfield{journal}{%
  \bibinfo {journal} {Phys. Rev. B}\ }%
  \textbf{\bibinfo {volume} {82}},\ \bibinfo {pages} {094409} (\bibinfo {year}
  {2010})%
  \bibAnnoteFile{NoStop}{LDB+10}%
\bibitem{Cla05}%
  \BibitemOpen
  \bibfield{author}{%
  \bibinfo {author} {\bibfnamefont{L.}~\bibnamefont{Claude}},\ }%
  \emph{\bibinfo {title} {Construction d'un microscope à effet tunnel à basse
  température et études d'impuretés magnétiques en surfaces}},\ \bibinfo
  {type} {{Ph.D.} thesis},\ \bibinfo {school} {\'{E}cole Polytechnique
  F\'{e}d\'{e}rale de Lausanne} (\bibinfo {year} {2005})%
  \bibAnnoteFile{NoStop}{Cla05}%
\bibitem{BH+09}%
  \BibitemOpen
  \bibfield{author}{%
  \bibinfo {author} {\bibfnamefont{P.}~\bibnamefont{B{\l}o\'{n}ski}}\ and\
  \bibinfo {author} {\bibfnamefont{J.}~\bibnamefont{Hafner}},\ }%
  \bibfield{journal}{%
  \bibinfo {journal} {J. Phys.: Condens. Matter}\ }%
  \textbf{\bibinfo {volume} {21}},\ \bibinfo {pages} {426001} (\bibinfo {year}
  {2009})%
  \bibAnnoteFile{NoStop}{BH+09}%
\bibitem{BST+09}%
  \BibitemOpen
  \bibfield{author}{%
  \bibinfo {author} {\bibfnamefont{T.}~\bibnamefont{Balashov}}, \bibinfo
  {author} {\bibfnamefont{T.}~\bibnamefont{Schuh}}, \bibinfo {author}
  {\bibfnamefont{A.~F.}\ \bibnamefont{Tak\'acs}}, \bibinfo {author}
  {\bibfnamefont{A.}~\bibnamefont{Ernst}}, \bibinfo {author}
  {\bibfnamefont{S.}~\bibnamefont{Ostanin}}, \bibinfo {author}
  {\bibfnamefont{J.}~\bibnamefont{Henk}}, \bibinfo {author}
  {\bibfnamefont{I.}~\bibnamefont{Mertig}}, \bibinfo {author}
  {\bibfnamefont{P.}~\bibnamefont{Bruno}}, \bibinfo {author}
  {\bibfnamefont{T.}~\bibnamefont{Miyamachi}}, \bibinfo {author}
  {\bibfnamefont{S.}~\bibnamefont{Suga}},\ and\ \bibinfo {author}
  {\bibfnamefont{W.}~\bibnamefont{Wulfhekel}},\ }%
  \bibfield{journal}{%
  \bibinfo {journal} {Phys. Rev. Lett.}\ }%
  \textbf{\bibinfo {volume} {102}},\ \bibinfo {pages} {257203} (\bibinfo {year}
  {2009})%
  \bibAnnoteFile{NoStop}{BST+09}%
\bibitem{EZWV08}%
  \BibitemOpen
  \bibfield{author}{%
  \bibinfo {author} {\bibfnamefont{C.}~\bibnamefont{Etz}}, \bibinfo {author}
  {\bibfnamefont{J.}~\bibnamefont{Zabloudil}}, \bibinfo {author}
  {\bibfnamefont{P.}~\bibnamefont{Weinberger}},\ and\ \bibinfo {author}
  {\bibfnamefont{E.~Y.}\ \bibnamefont{Vedmedenko}},\ }%
  \bibfield{journal}{%
  \bibinfo {journal} {Phys. Rev. B}\ }%
  \textbf{\bibinfo {volume} {77}},\ \bibinfo {pages} {184425} (\bibinfo {year}
  {2008})%
  \bibAnnoteFile{NoStop}{EZWV08}%
\bibitem{LSW03}%
  \BibitemOpen
  \bibfield{author}{%
  \bibinfo {author} {\bibfnamefont{B.}~\bibnamefont{Lazarovits}}, \bibinfo
  {author} {\bibfnamefont{L.}~\bibnamefont{Szunyogh}},\ and\ \bibinfo {author}
  {\bibfnamefont{P.}~\bibnamefont{Weinberger}},\ }%
  \bibfield{journal}{%
  \bibinfo {journal} {Phys. Rev. B}\ }%
  \textbf{\bibinfo {volume} {67}},\ \bibinfo {pages} {024415} (\bibinfo {year}
  {2003})%
  \bibAnnoteFile{NoStop}{LSW03}%
\bibitem{GRV+03}%
  \BibitemOpen
  \bibfield{author}{%
  \bibinfo {author} {\bibfnamefont{P.}~\bibnamefont{Gambardella}}, \bibinfo
  {author} {\bibfnamefont{S.}~\bibnamefont{Rusponi}}, \bibinfo {author}
  {\bibfnamefont{M.}~\bibnamefont{Veronese}}, \bibinfo {author}
  {\bibfnamefont{S.~S.}\ \bibnamefont{Dhesi}}, \bibinfo {author}
  {\bibfnamefont{C.}~\bibnamefont{Grazioli}}, \bibinfo {author}
  {\bibfnamefont{A.}~\bibnamefont{Dallmeyer}}, \bibinfo {author}
  {\bibfnamefont{I.}~\bibnamefont{Cabria}}, \bibinfo {author}
  {\bibfnamefont{R.}~\bibnamefont{Zeller}}, \bibinfo {author}
  {\bibfnamefont{P.~H.}\ \bibnamefont{Dederichs}}, \bibinfo {author}
  {\bibfnamefont{K.}~\bibnamefont{Kern}}, \bibinfo {author}
  {\bibfnamefont{C.}~\bibnamefont{Carbone}},\ and\ \bibinfo {author}
  {\bibfnamefont{H.}~\bibnamefont{Brune}},\ }%
  \bibfield{journal}{%
  \bibinfo {journal} {Science}\ }%
  \textbf{\bibinfo {volume} {300}},\ \bibinfo {pages} {1130} (\bibinfo {year}
  {2003})%
  \bibAnnoteFile{NoStop}{GRV+03}%
\bibitem{HKM+94}%
  \BibitemOpen
  \bibfield{author}{%
  \bibinfo {author} {\bibfnamefont{F.}~\bibnamefont{Huang}}, \bibinfo {author}
  {\bibfnamefont{M.~T.}\ \bibnamefont{Kief}}, \bibinfo {author}
  {\bibfnamefont{G.~J.}\ \bibnamefont{Mankey}},\ and\ \bibinfo {author}
  {\bibfnamefont{R.~F.}\ \bibnamefont{Willis}},\ }%
  \bibfield{journal}{%
  \bibinfo {journal} {Phys. Rev. B}\ }%
  \textbf{\bibinfo {volume} {49}},\ \bibinfo {pages} {3962} (\bibinfo {year}
  {1994})%
  \bibAnnoteFile{NoStop}{HKM+94}%
\bibitem{PSC+00}%
  \BibitemOpen
  \bibfield{author}{%
  \bibinfo {author} {\bibfnamefont{S.}~\bibnamefont{Padovani}}, \bibinfo
  {author} {\bibfnamefont{F.}~\bibnamefont{Scheurer}}, \bibinfo {author}
  {\bibfnamefont{I.}~\bibnamefont{Chado}},\ and\ \bibinfo {author}
  {\bibfnamefont{J.~P.}\ \bibnamefont{Bucher}},\ }%
  \bibfield{journal}{%
  \bibinfo {journal} {Phys. Rev. B}\ }%
  \textbf{\bibinfo {volume} {61}},\ \bibinfo {pages} {72} (\bibinfo {year}
  {2000})%
  \bibAnnoteFile{NoStop}{PSC+00}%
\bibitem{MBF+06}%
  \BibitemOpen
  \bibfield{author}{%
  \bibinfo {author} {\bibfnamefont{F.}~\bibnamefont{Meier}}, \bibinfo {author}
  {\bibfnamefont{K.}~\bibnamefont{von Bergmann}}, \bibinfo {author}
  {\bibfnamefont{P.}~\bibnamefont{Ferriani}}, \bibinfo {author}
  {\bibfnamefont{J.}~\bibnamefont{Wiebe}}, \bibinfo {author}
  {\bibfnamefont{M.}~\bibnamefont{Bode}}, \bibinfo {author}
  {\bibfnamefont{K.}~\bibnamefont{Hashimoto}}, \bibinfo {author}
  {\bibfnamefont{S.}~\bibnamefont{Heinze}},\ and\ \bibinfo {author}
  {\bibfnamefont{R.}~\bibnamefont{Wiesendanger}},\ }%
  \bibfield{journal}{%
  \bibinfo {journal} {Phys. Rev. B}\ }%
  \textbf{\bibinfo {volume} {74}},\ \bibinfo {pages} {195411} (\bibinfo {year}
  {2006})%
  \bibAnnoteFile{NoStop}{MBF+06}%
\bibitem{CFB+08}%
  \BibitemOpen
  \bibfield{author}{%
  \bibinfo {author} {\bibfnamefont{A.}~\bibnamefont{{Mosca Conte}}}, \bibinfo
  {author} {\bibfnamefont{S.}~\bibnamefont{Fabris}},\ and\ \bibinfo {author}
  {\bibfnamefont{S.}~\bibnamefont{Baroni}},\ }%
  \bibfield{journal}{%
  \bibinfo {journal} {Phys. Rev. B}\ }%
  \textbf{\bibinfo {volume} {78}},\ \bibinfo {pages} {014416} (\bibinfo {year}
  {2008})%
  \bibAnnoteFile{NoStop}{CFB+08}%
\bibitem{SBE+13}%
  \BibitemOpen
  \bibfield{author}{%
  \bibinfo {author} {\bibfnamefont{O.}~\bibnamefont{\v{S}ipr}}, \bibinfo
  {author} {\bibfnamefont{S.}~\bibnamefont{Bornemann}}, \bibinfo {author}
  {\bibfnamefont{H.}~\bibnamefont{Ebert}}, \bibinfo {author}
  {\bibfnamefont{S.}~\bibnamefont{Mankovsky}}, \bibinfo {author}
  {\bibfnamefont{J.}~\bibnamefont{Vack\'{a}\v{r}}},\ and\ \bibinfo {author}
  {\bibfnamefont{J.}~\bibnamefont{Min\'{a}r}},\ }%
  \bibfield{journal}{%
  \bibinfo {journal} {Phys. Rev. B}\ }%
  \textbf{\bibinfo {volume} {88}},\ \bibinfo {pages} {064411} (\bibinfo {month}
  {Aug}\ \bibinfo {year} {2013})%
  \bibAnnoteFile{NoStop}{SBE+13}%
\bibitem{WCF+97}%
  \BibitemOpen
  \bibfield{author}{%
  \bibinfo {author} {\bibfnamefont{R.}~\bibnamefont{Wu}}, \bibinfo {author}
  {\bibfnamefont{L.}~\bibnamefont{Chen}},\ and\ \bibinfo {author}
  {\bibfnamefont{A.~J.}\ \bibnamefont{Freeman}},\ }%
  \bibfield{journal}{%
  \bibinfo {journal} {J. Magn. Magn. Materials}\ }%
  \textbf{\bibinfo {volume} {170}},\ \bibinfo {pages} {103 } (\bibinfo {year}
  {1997})%
  \bibAnnoteFile{NoStop}{WCF+97}%
\bibitem{THO+07}%
  \BibitemOpen
  \bibfield{author}{%
  \bibinfo {author} {\bibfnamefont{M.}~\bibnamefont{Tsujikawa}}, \bibinfo
  {author} {\bibfnamefont{A.}~\bibnamefont{Hosokawa}},\ and\ \bibinfo {author}
  {\bibfnamefont{T.}~\bibnamefont{Oda}},\ }%
  \bibfield{journal}{%
  \bibinfo {journal} {J. Phys.: Condens. Matter}\ }%
  \textbf{\bibinfo {volume} {19}},\ \bibinfo {pages} {365208} (\bibinfo {year}
  {2007})%
  \bibAnnoteFile{NoStop}{THO+07}%
\bibitem{NCZ+01}%
  \BibitemOpen
  \bibfield{author}{%
  \bibinfo {author} {\bibfnamefont{B.}~\bibnamefont{Nonas}}, \bibinfo {author}
  {\bibfnamefont{I.}~\bibnamefont{Cabria}}, \bibinfo {author}
  {\bibfnamefont{R.}~\bibnamefont{Zeller}}, \bibinfo {author}
  {\bibfnamefont{P.~H.}\ \bibnamefont{Dederichs}}, \bibinfo {author}
  {\bibfnamefont{T.}~\bibnamefont{Huhne}},\ and\ \bibinfo {author}
  {\bibfnamefont{H.}~\bibnamefont{Ebert}},\ }%
  \bibfield{journal}{%
  \bibinfo {journal} {Phys. Rev. Lett.}\ }%
  \textbf{\bibinfo {volume} {86}},\ \bibinfo {pages} {2146} (\bibinfo {year}
  {2001})%
  \bibAnnoteFile{NoStop}{NCZ+01}%
\bibitem{CS65}%
  \BibitemOpen
  \bibfield{author}{%
  \bibinfo {author} {\bibfnamefont{J.}~\bibnamefont{Crangle}}\ and\ \bibinfo
  {author} {\bibfnamefont{W.~R.}\ \bibnamefont{Scott}},\ }%
  \bibfield{journal}{%
  \bibinfo {journal} {J. Appl. Phys.}\ }%
  \textbf{\bibinfo {volume} {36}},\ \bibinfo {pages} {921} (\bibinfo {year}
  {1965})%
  \bibAnnoteFile{NoStop}{CS65}%
\bibitem{Zel93}%
  \BibitemOpen
  \bibfield{author}{%
  \bibinfo {author} {\bibfnamefont{R.}~\bibnamefont{Zeller}},\ }%
  \bibfield{journal}{%
  \bibinfo {journal} {Modelling Simul. Mater. Sci. Eng.}\ }%
  \textbf{\bibinfo {volume} {1}},\ \bibinfo {pages} {553} (\bibinfo {year}
  {1993})%
  \bibAnnoteFile{NoStop}{Zel93}%
\bibitem{SMM+08}%
  \BibitemOpen
  \bibfield{author}{%
  \bibinfo {author} {\bibfnamefont{O.}~\bibnamefont{\v{S}ipr}}, \bibinfo
  {author} {\bibfnamefont{J.}~\bibnamefont{Min\'{a}r}}, \bibinfo {author}
  {\bibfnamefont{S.}~\bibnamefont{Mankovsky}},\ and\ \bibinfo {author}
  {\bibfnamefont{H.}~\bibnamefont{Ebert}},\ }%
  \bibfield{journal}{%
  \bibinfo {journal} {Phys. Rev. B}\ }%
  \textbf{\bibinfo {volume} {78}},\ \bibinfo {pages} {144403} (\bibinfo {year}
  {2008})%
  \bibAnnoteFile{NoStop}{SMM+08}%
\end{thebibliography}

% File *.bbl inserted manually in order to avoid need for BibTeX
% cooperation;  this is an aid to PR publishing

%Merlin.mbs v4.21 2009-07-09.
%

\end{document}